\documentclass[a4paper,11pt]{article}
\pdfoutput=1 
\usepackage{jcappub} 
\usepackage[T1]{fontenc} 
\begin{document}

\title{Anisotropic stars in modified gravity: An extended gravitational decoupling approach}                      

\author[a]{S. K. Maurya,}
\author[c]{B. Mishra,}
\author[d]{Saibal Ray,}
\author[a]{Riju Nag}

\affiliation[a]{Department of Mathematics and Physical Science, College of Arts and Science, University of Nizwa, Nizwa, Sultanate of Oman}
\affiliation[c]{Department of Mathematics, Birla Institute of Technology and Science-Pilani, Hyderabad Campus, Hyderabad 500078, India}
\affiliation[d]{Department of Physics, Government College of Engineering and Ceramic Technology, Kolkata 700010, West Bengal, India}
\affiliation[a]{Department of Mathematics and Physical Science, College of Arts and Science, University of Nizwa, Nizwa, Sultanate of Oman}

\emailAdd{sunil@unizwa.edu.om}
\emailAdd{bivu@hyderabad.bits-pilani.ac.in}
\emailAdd{saibal@associates.iucaa.in}
\emailAdd{rijunag@gmail.com}

\date{\today}

\abstract{In the work we present investigation on decoupling gravitational sources under the framework of $f(R,T)$ gravity. Basically the complete geometric deformation technique has been employed here which facilitates finding exact solutions to the anisotropic astrophysical system smoothly without imposing any particular ansatz for deformation function. Along with this we have also used 5-dimensional Euclidean spacetime in order to describe the embedding Class I spacetime for getting a solvable spherical physical system.  The solutions thus obtained show physically interesting as well as viable with new possibilities to sought for. Especially, from the present investigation it is worthy to note that the mixture $f(R,T)$ + CGD translate the scenario beyond the pure GR realm and hence clearly helps to enhance the features of the interior astrophysical aspects of compact stellar objects. Therefore to check the physical acceptability and stability of the stellar system based on the obtained solutions, we have performed a few physical tests which satisfy all the stability criteria, including nonsingular nature of the density as well as pressure.}


\maketitle
\flushbottom

\section{Introduction}
The modification in the gravitational sector has been proposed from time to time because of the correction in the gravitation action. These corrections have become inevitable, when already existing gravitational theories unable to address certain key issues of the present universe. There are several such corrections incorporated in recent past such as, for cosmological applications \cite{Gurevich79,Whitt84}, string theory \cite{Hoissen85,Tsevtlin87}, teleparallel gravity \cite{Aldrovandi12}, unimodular gravity \cite{Nojiri16}, $f(R)$ gravity \cite{Nojiri07,Nojiri11}, $f(R,G)$ gravity \cite{Nojiri06,Bazeia07}, $f(Q,T)$ gravity \cite{Xu19} and so on. Several cosmological observations  have confirmed the late time cosmic acceleration phenomena and this development has changed our understanding on the universe. This recent development has been led to new concepts and ideas. The late time cosmic acceleration phenomena attributed to an exotic form of energy called the dark energy and general relativity (GR) has certain limitations in addressing this issue. There are different exotic matter fields that simulate negative pressure and positive energy density to explain this bizarre fact, but a suitable geometrical modification without adding any exotic matter possibly settle this issue. Harko et al. \cite{Harko11} proposed the $f(R,T)$ gravity by assuming a weak coupling between matter and geometry. The geometrical part of Einstein-Hilbert action has been modified by assuming the arbitrary function $f(R,T)$, where $R$ and $T$ respectively denote the Ricci scalar and the trace of the energy momentum tensor. The trace incorporated in this function may associate with the existence of exotic imperfect fluids. Three form of the functions are suggested as: (i) $f(R,T)=R+2f(T)$, (ii) $f(R,T)=f_1(R)+f_2(T)$ and (iii) $f(R,T)=f_1(R)+f_2(R)f_3(T)$, where $f_1(R)$, $f_2(R)$, $f(T)$, $f_2(T)$ and $f_3(T)$ are arbitrary functions of their respective arguments. The first case can be reduced to GR under certain condition and this has been widely used to address issues related to cosmology and astrophysics. 

In recent time, $f(R,T)$ gravity theory has been of great interest and been used widely in literature to address many issues related to astrophysics and cosmology. The issue of late time cosmic dynamics and the anisotropy behaviour of the expansion has been partially addressed in $f(R,T)$ gravity. Some of the key findings in $f(R,T)$ gravity are given here. Alvarenga et al. \cite{Alvarenga13}  have studied, in metric formalism, the evolution of scalar cosmological perturbations. Balakin and Bochkarev \cite{Balakin13} have investigated the rip cosmology where as Noureen et al. \cite{Noureen15} have shown the implications of shear-free condition on the instability range of an anisotropic fluid. Baffou et al. \cite{Baffou15} have solved the cosmological evolution of the cosmological parameters numerically in $f(R,T)$ gravity. Mishra et al. \cite{Mishra19} have introduced the hybrid scale factor to study the dynamical behaviour of the cosmological model of the universe and also introduced the squared trace model \cite{Mishra20} in $f(R,T)$ gravity. In order to address the singularity issue, bouncing cosmology has also been studied in $f(R,T)$ gravity. Shabani and Ziaie \cite{Shabani18} have introduced the effective fluid by defining the effective energy density and pressure to present the bouncing cosmological model. Tripathy et al. \cite{Tripathy21} have studied the bouncing cosmology in $f(R,T)$ gravity and performed the stability analysis of the models under linear homogeneous perturbations solved the bouncing. The $f(R,T)$ gravity theory has also been significant in the study of wormhole solutions. Zubair et al. \cite{Zubair16} have obtained the wormhole solutions in $f(R,T)$ gravity and shown that the wormhole can be constructed without exotic matter in few regions of space-time. Elizalde and Khurshudyan \cite{Elizalde19} have obtained the wormhole solution and explored the observational possibilities for testing these models. Yousaf et al. \cite{Yousaf16} have investigated the irregularity factors of self-gravitating spherical star that evolves in the presence of an imperfect fluid. Das et al. \cite{Das17} have studied the gravastar in $f(R,T)$ gravity and obtained a set of singularity-free and exact solution of the gravastar. Abbas et al. \cite{Abbas19} have studied the charged perfect fluid spherically symmetric gravitational collapse and commented that the singularity is formed earlier than the apparent horizons. 

In the context of GR, the charged compact objects  have been investigated with the interpretation of anisotropic system from Einstein-Maxwell field equations. The anisotropic factor has been mimicked from the electric field intensity and the model with this assumption helps in obtaining the stability of static fluid in presence of charge. The solution of Einstein-Maxwell equations has been instrumental in describing the astrophysical compact objects \cite{Mehra80,Bonnor89,Ivanov02,Ray03}. However, since last one decade the researchers have given attention to construct charged compact star model in $f(R,T)$ gravity to understand its behaviour in the geometrically extended gravity. Moraes et al. \cite{Moraes16} have studied the stellar equilibrium configurations of compact stars in $f(R,T)$ gravity. Islam and Basu \cite{Islam18} have constructed model of a compact star in presence of magnetic fluid and suggested that these solutions will enable to describe the interior of compact objects. With the Krori-Barua solutions and in an anisotropic distribution, Sharif and Waseem \cite{Sharif19a} have analysed the effects of charge on the nature of relativistic compact star candidates. In the scope of this extended gravity, Yadav et al. \cite{Yadav20} have proposed the existence of non-exotic compact star, that validates the energy conditions and stability of the model. Biswas et al. \cite{Biswas20} have studied the anisotropic spherically symmetric strange star and have shown the validity of the conditions used in the model. Maurya et al. \cite{Maurya19} have suggested an embedded approach to study the existence of compact structures that describes  anisotropic matter distributions in the framework of matter geometry coupling. Also, Maurya and Tello-Ortiz \cite{Maurya20} have extended the isotropic Durgapal-Fuloria model and investigated the high dense charged anisotropic compact structure in an isotropic background. Rahaman et al. \cite{Rahaman20} have predicted the exact redii of the values of the coupling parameter involve in $f(R,T)$ gravity by considering the observed mass values of six compact stars.  Rej et al. \cite{Rej21} have obtained singularity free model of charged anisotropic compact star in $f(R,T)$ gravity.

The standard model is based on the homogeneity of space and large scale isotropy, however a small scale of anisotropy can be expected in the universe \cite{Picon04,Jaffe05,Buiny06,Watanabe09,Saadeh16,Mishra19}. The isotropy and homogeneity can be observed through the space-time under consideration. But, the space-time of compact objects lead to the anisotropic features with the inhomogeneous matter distribution. The difference between the radial and tangential component pressure leads to the anisotropic pressure. Due to this anisotropy feature, the physical properties like gravitational redshift, energy density, total mass etc. are affected. There are many causes behind the anisotropy, e.g. pion condensation \cite{Sawyer72}, phase transitions \cite{Sokolov80}, immense magnetic field of neutron star \cite{Weber99}, strong electric field \cite{Usov04} and refs. therein \cite{Schunck03,Mak03,Rahaman10}. In addition to these, the form of gravitational tidal effects believed to be another reason of anisotropy in the compact star and this has been responsible for the deformation \cite{Doneva12,Biswas19,Rahmansyah20,Raoupas20,Das2021a,Das2021b}. 

A literature survey shows that now-a-days, several modified/extended theories of gravity are being used to investigate compact stellar models. Therefore, it is very much needed to understand the inherent geometry of the spacetime and the process to embed a 4D space-time. It is known that Karmarkar \cite{Karmarkar48} has embedded 4-dimensional spacetime into 5-dimensional Euclidean space. This embedding actually simplifies the process of solving the Einstein field equations. Further, different types of manifolds are linked by embedding 4-dimensional Einstein field equations into 5-dimensional flat spacetime \cite{Rippl95,Lidsey97}. It is notable that in general relativistic background Maurya et al. \cite{Maurya16} have obtained the exact generalized model for anisotropic compact stars of embedding Class I and tested the viability of the model by performing different physical tests, viz. the energy conditions, stability analysis and mass-radius relation. It is also to note that Salako et al. \cite{Salako20} have studied the existence of strange starts in $f(\mathcal{T},T)$ gravity, where $\mathcal{T}$ be the torsion tensor. Waheed et al. \cite{Waheed20} have used the Karmarkar condition~\cite{Karmarkar48} to find the physically acceptable solution for compact star in $f(R,T)$ gravity. In $f(R,T)$ gravity, Ahmed and Abbas \cite{Ahmed20} have studied the gravitational collapse using the Karmarkar condition \cite{Karmarkar48} to the spherically symmetric non-static radiating star. 

Under the above discussion we would like to now explain here our technique to adopt for solving the Einstein field equations in the anisotropic domain as efficient as possible. The newly adopted method which is known in the literature as the gravitational decoupling (GD) approach, that allows one to decouple the Einstein equations \cite{Heras2018,Ovalle2018a,Ortiz2020,Meert2021} \footnote{{A detailed discussion on the origin as well as feature of GD can be obtained in Ref. \cite{Ovalle2019a,Maurya2020,Ovalle2020a,Ovalle2020b,Maurya2021}}}. This gravitational decoupling approach is applied in system via minimal geometric deformation (MGD) and its extension, called extended MGD or complete geometric deformation (CGD). These efficient technique have been employed by many investigators to tackle the system \cite{Ovalle2017,Ovalle2018b,Ovalle2018c,Ovalle2018d,Estrada2018,Gabbanelli2018,Morales2018a,Morales2018b,Ovalle2019b,Estrada2019,Gabbanelli2019,Torres2019,Rincon2019,Contreras2019b,Rocha2020,Sharif2018,Sharif2020a,Sharif2020b,Ovalle2021,Heras2021}. Here, the extended MGD or complete geometric deformation (CGD) which basically has been used by us in the present paper to investigate the exact solution for compact star in modified gravity theory without imposing any deformation function. The systematic approach of this technique has been explained in details later on in a forgoing section.  

The paper is organized as follows: In Section 2, some preliminaries of the $f(R,T)$ gravity for gravitational decoupling along with the Einstein's field equations and the Class I condition are discussed. In Section 3, we have adopted a systematic procedure to solve the field equations for decoupled system. In Section 4, the matching conditions for the astrophysical system has been elaborated. The physical analysis of the problem are done in Section 5 and in Section 6 discussions and conclusions are presented.

\section{Preliminaries of the mathematical frameworks} 

\subsection{Generalized $f(R,T)$  gravity for gravitational decoupling system}\label{sec2}

The generalized integral action for the $f(R,T)$ formulation, one can write by adding extra source as~\cite{Ovalle2019a}
\begin{eqnarray} \label{eq1}
S=\frac{1}{16\pi}\int f(R,{T})\sqrt{-g} d^{4}x+\int {L}_{m}\sqrt{-g}d^{4}x +\beta\int {L}_{X}\sqrt{-g} d^{4}x,
\end{eqnarray}
where ${L}_{m}$ denotes the matter Lagrangian, $\beta$ denotes a coupling constant and ${L}_{X}$ be the Lagrangian density of a new sector. Here ${L}_{X}$ is not required to be essentially defined by GR, however this can create the alterations in GR as argued in~\cite{Ovalle2019a}.   

Let us now vary the action with respect to the metric tensor $g^{\mu\nu}$ which yields the field equations as follows 
\begin{eqnarray} \label{eq2}
\left( R_{\mu\nu}- \nabla_{\mu} \nabla_{\nu} \right)f_R (R,{T}) +\Box f_R (R,{T})g_{\mu\nu} - \frac{1}{2} f(R,{T})g_{\mu\nu} \nonumber \\ = 8\pi\left( T_{\mu\nu} +\beta \theta_{\mu\nu}\right) - f_{T}(R,{T})\, \left(T_{\mu\nu}  +\Theta_{\mu\nu}\right),
\end{eqnarray}
where $f_{R}(R,T)=\partial f(R,T)/\partial R$ and $f_{T}(R,T)=\partial f(R,T)/\partial T$, respectively. 

Again, the energy-momentum tensor (EMT) $T_{\mu\nu}$ and extra source $\theta_{\mu\nu}$, along with $\Theta_{\mu\nu}$ can be provided as
\begin{eqnarray}\label{eq3}
T_{\mu\nu}&=&g_{\mu\nu}{L}_m-2\partial{L}_m/\partial g^{\mu\nu},\\ \label{eq4}
\theta_{\mu\nu}&=&-g_{\mu\nu}{L}_{X}+2\partial{L}_{X}/\partial g^{\mu\nu},\\ \label{eq5}
{\Theta_{\mu\nu}}&=& {g^{\gamma\,\epsilon}\delta T_{\gamma\,\epsilon}/\delta g^{\mu\nu}},
\end{eqnarray}
where the conceptual meaning of this special tensor $\Theta_{\mu\nu}$ will be  physically meaningful later on.

Hence the Einstein tensor $G_{\mu\nu}$, after rearranging of Eq. (\ref{eq2}), can be given as
\begin{eqnarray}\label{eq6}
&& G_{\mu\nu}=\frac{1}{f_{R}\left(R,{T}\right)}\big[8\pi\left(T_{\mu\nu}-\beta \theta_{\mu\nu}\right)+\frac{1}{2}\left(f\left(R,{T}\right)-Rf_{R}\left(R,{T}\right)\right)g_{\mu\nu} \nonumber\\ &&  -\left(T_{\mu\nu}+\Theta_{\mu\nu}\right)f_{{T}}\left(R,{T}\right) -\left(g_{\mu\nu}\Box-\nabla_{\mu}\nabla_{\nu}\right)f_{R}\left(R,{T}\right)+8\,\pi\,E_{\mu\nu}\big].
\end{eqnarray}

The respective conservation equation now leads to
\begin{eqnarray}\label{eq7}
&& \nabla^{\mu}T_{\mu\nu}=\frac{f_{T}(R,{T})}{8\pi -f_{T}(R,{T})}\bigg[(T_{\mu\nu} +\Theta_{\mu\nu})\nabla^{\mu}  \ln f_{T}(R,{T})  \nonumber\\ && +\nabla^{\mu}\Theta_{\mu\nu}-\frac{1}{2}g_{\mu\nu}\nabla^{\mu}{T} \frac{8\pi}{f_{{T}}\left(R,{T}\right)} \big(\beta\,\nabla^{\mu}\theta_{\mu\nu}\big)\bigg].
\end{eqnarray}

The energy-momentum tensor $T_{\mu\nu}$ for the anisotropic matter distribution can be taken as
\begin{equation}\label{eq8}
{T}_{\mu\nu}=\left({\rho}+{p}_t\right)u_{\mu}u_{\nu}+{p}_t g_{\mu\nu}+(p_r-p_t)\,\zeta_{\mu}\zeta_{\nu},
\end{equation}
where ${u^{\mu}}$ is the four velocity, satisfying $u_{\mu}u^{\mu}=-1$ and $u_{\mu}\nabla^{\nu}u_{\mu}=0$ while $\rho$, $p_r$ and $p_t$ are respectively the matter density, radial pressure and tangential pressure of the system.

Here, the matter Lagrangian considered as, ${L}_m=-\mathcal{P}$, where  $\mathcal{P}=-\frac{1}{3}\big(p_r+2\,p_t)$. It is worthy to mention here that the matter Lagrangian appears in the field equations (\ref{eq6}) and the conservation equations (\ref{eq7}) can be realized according to its choice \footnote{{See \cite{new2} and references therein for a modern, recent and detailed discussion about this subject in the framework of modified gravity theories, included $f(R,T)$ gravity.}}. However, in the context of GR, this choice would not affect the observational outcomes. So, when the matter field is described a perfect fluid, it is quite relevant to choose a particular form of the matter Lagrangian. The choice we mention here is consistent since it provides well-established Lagrangian density. Moreover, in the limit $\chi\rightarrow 0$, it reduces to the perfect fluid distribution in the context of GR \cite{new1}.

So, from Eq. (\ref{eq3}) one gets
\begin{equation}
\frac{\delta\,T_{\mu \nu}}{\delta\,g^{\gamma \epsilon}}=\left(\frac{\delta\,g_{\mu \nu}}{\delta\,g^{\gamma \epsilon}}\right)\,L_{m}+g_{\mu \nu}\,\left(\frac{\partial\,L_{m}}{\partial\,g^{\gamma \epsilon}}\right)-2\,\frac{\partial^{2}\,L_{m}}{\partial\,g^{\mu \nu}\,\partial\,g^{\gamma \epsilon}}.    \label{eq9}
\end{equation}

Now, using $\delta\,g_{\mu \nu}/\delta\,g^{\gamma \epsilon}=-g_{\mu\sigma}g_{\nu \chi}\delta^{\sigma \chi}_{\gamma \epsilon}$, the above equation becomes
\begin{eqnarray}
&&\hspace{-0.5cm}\frac{\delta\,T_{\mu \nu}}{\delta\,g^{\gamma \epsilon}}=g_{\mu \nu}\,\left(\frac{\partial\,L_{m}}{\partial\,g^{\gamma \epsilon}}\right)-g_{\mu\sigma}\,g_{\nu \chi}\,\delta^{\sigma \chi}_{\gamma \epsilon}\,L_{m}-2\,\frac{\partial^{2}\,L_{m}}{\partial\,g^{\mu \nu}\,\partial\,g^{\gamma \epsilon}}.~~~\label{eq10}   
\end{eqnarray}

Plugging (\ref{eq10}) in (\ref{eq5}), we obtain
\begin{equation}\label{eq11}
\Theta_{\mu\nu}=-2\,T_{\mu\nu}+g_{\mu\nu}\,L_{m}- 2\,g^{\gamma \epsilon}\frac{\partial^{2}\,L_{m}}{\partial\,g^{\gamma \epsilon}\,\partial\,g^{\mu\nu}}.   
\end{equation}

Again, employing (\ref{eq3}) and ${L}_m=-\mathcal{P}$, we finally get
\begin{equation}\label{eq12}
\Theta_{\mu\nu}=-2\,T_{\mu\nu}-\mathcal{P}\,g_{\mu\nu}.    
\end{equation}

Now, taking into account the work of Harko et al.~\cite{Harko11}, we consider the linear form of $f(R,T)$ as
\begin{eqnarray}\label{eq13}
f(R,{T})=R+2\chi{T}, 
\end{eqnarray}
where $\chi$ is dimensionless and known as the coupling constant. The linear form of $f(R,T)$ is quite successful in the context of astrophysical and cosmological models. We have elaborated its successes in the Introduction. 

By substituting the $f(R,T)$ functional (\ref{eq13}) in Eq. (\ref{eq6}) we find
\begin{eqnarray} \label{eq14}
&& \hspace{-0.2cm}
G_{\mu\nu}=8\pi \left(T_{\mu\nu}-\beta\,\theta_{\mu\nu}\right) +\chi(2T_{\mu\nu}+2\mathcal{P} g_{\mu\nu}+{T}g_{\mu\nu})  \nonumber\\ && \hspace{2.2cm} =8\pi\,\big( T_{\mu\nu}-\beta\,\theta_{\mu\nu}+\hat{T}_{\mu\nu}\big),
\end{eqnarray}
where we denote $\hat{T}_{\mu\nu}$ as
\begin{eqnarray}\label{eq15}
\hat{T}_{\mu\nu}=\frac{\chi}{8\,\pi}\,(2T_{\mu\nu}+2\,\mathcal{P} g_{\mu\nu}+{T}g_{\mu\nu}),
\end{eqnarray}
with ${T}_{\mu\nu}$ as given by Eq. (\ref{eq8}). 

Hence, from the conservation of (\ref{eq14}) one obtains
\begin{eqnarray}\label{eq16}
&&\hspace{0.5cm}\nabla^{\mu}\big( T_{\mu\nu}-\beta\,\theta_{\mu\nu}+\hat{T}_{\mu\nu}\big)=0. \end{eqnarray}

At this stage we would like to mention the terms $\mathcal{P}$ and $T$ as
\begin{eqnarray}
\mathbf{\mathcal{P}}=-\frac{1}{3} (p_r+2\,p_t)~~~\text{and} ~~~T=-\rho+p_r+2\,p_t. \nonumber
\end{eqnarray}

\subsection{The Einstein field equations for the decoupled system}\label{sec3}
 We consider the most general line element to describe a spherically symmetric and static spacetime which is given by
\begin{equation}\label{eq17}
ds^{2} =- e^{y(r) } \, dt^{2}+e^{\lambda(r)} dr^{2} +r^{2}(d\theta ^{2} +\sin ^{2} \theta \, d\phi ^{2}),
\end{equation}
where the metric potential $y$ and $\lambda$ are function of the radial coordinate $r$ only, i.e, $y=y(r)$ and $\lambda=\lambda(r)$.
Let us write the non-zero components of the field equations under the static spherically symmetric line element (\ref{eq17}) can be provided as 
\begin{eqnarray}\label{eq18}
&&\hspace{-0.6cm} \frac{e^{-\lambda}}{8\pi}\left(-\frac{1}{r^2}+\frac{\lambda^{\prime}}{r}+\frac{e^{\lambda}}{r^2}\right)={\rho}-\hat{T}^{0}_{\ 0}+\beta\, \theta^{0}_{\ 0} , \\\label{eq19}
&&\hspace{-0.6cm} \frac{e^{-\lambda}}{8\pi}\left(\frac{1}{r^2}+\frac{y^{\prime}}{r}-\frac{e^{\lambda}}{r^2}\right)={p_r}+\hat{T}^{1}_{\ 1}-\beta\, \theta^{ 1}_{\ 1}, \\\label{eq20}
&&\hspace{-0.6cm} \frac{e^{-\lambda}}{32\pi}\left(2y^{\prime\prime}+y^{\prime2}-\lambda^{\prime}y^{\prime}+2\frac{y^{\prime}-\lambda^{\prime}}{r}\right)={p_t}+\hat{T}^{2}_{\ 2}-\beta\, \theta^{2}_{\ 2}.~~~
\end{eqnarray}

In the above a `prime' as usual denotes differentiation with respect to the radial coordinate $r$. Moreover, $\hat{T}^{0}_{\ 0}$ and $\hat{T}^{1}_{\ 1}$ are expressed as
\begin{eqnarray}\label{eq21}
&& \hat{T}^{0}_{0}=\frac{\chi}{24\,\pi}(-9\rho+p_r+2p_t),\\\label{eq22}
&&\hat{T}^{1}_{\ 1}=\frac{\chi}{24\,\pi}(-3\rho+7p_r+2p_t),\\\label{eq23}
&& \hat{T}^{2}_{\ 2}=\frac{\chi}{24\,\pi}(-3\rho+p_r+8p_t). 
\end{eqnarray}

\subsection{The Class I condition for the decoupled system}
In general, the 4-dimensional spherically symmetric spacetime given by the metric (\ref{eq17}), describes a spacetime of Class II. This shows that it is required a 6-dimensional pseudo-Euclidean space for embedding. In this connection, Gupta et al.~\cite{Gupta} have provided 6-dimensional Euclidean spacetime in the form
\begin{eqnarray}\label{eq49}
 ds^2=- d\mathbb{Y}_1^2- d\mathbb{Y}_2^2- d\mathbb{Y}_3^2+ d \mathbb{X}_1^2+d \mathbb{X}_2^2\pm d \mathbb{X}_3^2, 
\end{eqnarray}
with the particular transformation as follows:
\begin{eqnarray}
&&\hspace{-0.3cm} \mathbb{Y}_1=r\,\sin\theta \cos\phi, ~\mathbb{Y}_2=r\,\sin\theta \sin\phi,~\mathbb{Y}_3=r\,\cos\theta,~\nonumber\\&&\hspace{-0.3cm}  \mathbb{X}_1=K\,e^{y(r)/2}\,\cosh\Big(\frac{t}{K}\Big),~\mathbb{X}_2=K\,e^{y(r)/2}\,\sinh\Big(\frac{t}{K}\Big),~ \mathbb{X}_3=\mathbb{Z}(r). \nonumber
\end{eqnarray}

Then Eq. (\ref{eq49}) readily takes the form
\begin{eqnarray} \label{eq50}
 ds^2=e^{y(r)/2}\,dt^2- \Big( 1+ \frac{K^2\,y^{\prime2}(r)\,e^{y(r)}}{4}\,\pm \mathbb{Z}^{\prime}(r) \Big) dr^2 -r^2(d\theta^2+\sin^2\theta\,d\phi^2).
\end{eqnarray}

After comparing from Eqs. (\ref{eq3}) and (\ref{eq50}), we have
\begin{eqnarray}\label{eq51}
 e^{\lambda(r)}=\Big( 1+ \frac{K^2\,y^{\prime2}(r)\,e^{y(r)}}{4}\,\pm \mathbb{Z}^{\prime}(r) \Big)
\end{eqnarray}  
and
\begin{eqnarray}\label{eq52}
 ds^2=- d\mathbb{Y}_1^2- d\mathbb{Y}_2^2- d\mathbb{Y}_3^2+ d \mathbb{X}_1^2+d \mathbb{X}_2^2+d \mathbb{X}_3^2.
\end{eqnarray}

On the other hand, the Karmarkar condition~\cite{Karmarkar48} proposes that for any spherically symmetric spacetime (static as well as non-static) to be a Class I, the spacetime must satisfy the following condition 
\begin{eqnarray}\label{eq53}
R_{1010}R_{2323} =R_{1212}R_{3030} + R_{2102}R_{3103}, 
\end{eqnarray}
subject to $R_{2323}\neq 0$~\cite{Pandey1982}, where the quantities regarding the Riemann components for the metric (\ref{eq17}) are given as
\begin{eqnarray}\label{eq54}
&&\hspace{-0.3cm} R_{2323}=-r^2\,(1-e^{-\lambda})\, \sin^2\theta,~~R_{1212}=\frac{\lambda' r}{2},~~ R_{3103}=0,\nonumber \\
&&\hspace{-0.3cm} R_{1010}=-e^y \left[\frac{y\,''}{2}+\frac{y\,'^2}{4}-\frac{\lambda' y\,'}{4}\right],~~R_{2102}=0,\nonumber\\&& 
\hspace{-0.3cm}R_{3030}=-\sin^2{\theta} \frac{y' r e^{y-\lambda}}{2}.
\end{eqnarray}

After inserting the Riemann components in condition (\ref{eq15}), we obtain  
\begin{eqnarray}\label{eq55}
&& 2\frac{y^{\prime\prime}}{y^{\prime}}+y^{\prime}=\frac{\lambda^{\prime}e^{\lambda}}{e^{\lambda}-1},
\end{eqnarray}
with $e^{\lambda}\neq 1$. 

The solution of the above differential equation requires for the space time (\ref{eq17}) to be a Class I. Hence the integration of (\ref{eq16}) provides the relation between the gravitational potentials   
\begin{eqnarray}\label{eq56}
 e^{\lambda}=1+A\,y^{\prime 2}(r)\,e^{y(r)},
\end{eqnarray}
where $A$ is the integration constants. 

From (\ref{eq51}) and (\ref{eq56}), we note that the transformation given below (\ref{eq49}) give 5-dimensional Euclidean spacetime in order to describe the embedding  Class I spacetime if and only if $\mathbb{Z}=0$ and $A=\frac{K^2}{4}$.

\section{A systematic procedure for solving the field equations for decoupled system via CGD technique} 
A close observation on the field equations (\ref{eq18})--(\ref{eq20}) clearly shows that a closed exact solution is not a easy and trivial task. Therefore, we employ the complete geometric deformation (CGD) technique for solving these system of equations in a unique way. This CGD technique provides a systematic approach which are: first, split the decoupled system into two subsystems, and secondly solve these system individually. 

\subsection{Splitting the decoupled system via CGD approach}
In this approach, we basically deform the gravitational potentials ${\lambda(r)}$  and   ${y(r)}$ over a linear transformation given as
\begin{eqnarray}\label{eq24}
\lambda(r)&\mapsto& -\ln[\xi(r)+\beta h(r)],\\\label{eq25}
y(r)&\mapsto& \eta(r)+\beta g(r), 
\end{eqnarray}
where $h(r)$ and $g(r)$ denote the decoupling functions corresponding to the radial and the temporal components of the line element~(\ref{eq17}). Here we take the deformation along with both the radial component and the temporal components, i.e. $f(r)\ne 0$ and $g(r) \ne 0$ so that it is known the complete geometric deformation (CGD). Also, It is always possible to separate the new piece $\theta_{\mu\nu}$ from the seed matter sector for pure $f(R,T)$ system.  

Now, inserting Eqs. (\ref{eq24}) and (\ref{eq25}) into the system of equations (\ref{eq18})--(\ref{eq20}) we can have
\begin{eqnarray}\label{eq27}
&&\hspace{-0.8cm} {8\,\pi}\,\big(\rho-\hat{T}^{0}_{\ 0}\big)+8\pi\beta\theta^{0}_{\ 0} =\bigg[\frac{1}{r^{2}}-\frac{\xi}{r^{2}}-\frac{\xi^{\prime}}{r}\bigg]
-\beta\bigg[\frac{h}{r^{2}}+\frac{h^{\prime}}{r}\bigg],~~~~ \\\label{eq28}
&&\hspace{-0.8cm} {8\,\pi}\,\big(p_r +\hat{T}^1_1\big)-8\pi\beta\theta^{1}_{\ 1}= \bigg[\xi\left(\frac{1}{r^{2}}+\frac{\eta^{\prime}}{r}\right)-\frac{1}{r^{2}}\bigg] +\beta h\bigg[\frac{1}{r^{2}}+\frac{\eta^{\prime}}{r}\bigg]+\beta \frac{\xi\,g^\prime}{r},\\\label{eq29}
&&\hspace{-0.8cm} {8\,\pi}\,\big(p_t +\hat{T}^2_2\big)-8\pi\beta\theta^{2}_{\ 2}= \bigg[\frac{\xi}{4}\left(2\eta^{\prime\prime}+\eta^{\prime2}+2\frac{\eta^{\prime}}{r}\right)+\frac{\xi^{\prime}}{4}\left(\eta^{\prime}+\frac{2}{r}\right)\bigg]\nonumber\\&&\hspace{3.cm}+\beta\bigg[\frac{h}{4}\left(2\eta^{\prime\prime}+\eta^{\prime2}+2\frac{\eta^{\prime}}{r}\right)
+\frac{h^{\prime}}{4}\left(\eta^{\prime}+\frac{2}{r}\right)+\Psi(r)\bigg],
\end{eqnarray}
where
\begin{eqnarray}\label{eq30}
\Psi(r)=\frac{\mu ^\prime h^\prime}{4}+ \frac{\mu }{4}\Big(2h^{\prime \prime}~+\beta h^{\prime 2}~+ \frac{2h^\prime}{r}~+2 \xi ^\prime h^\prime \Big)
\end{eqnarray}
where $\Psi(r)$ is given by the Eq. (\ref{eq30}).  
\begin{eqnarray} \label{eq38a}
 m(r)=\underbrace{\frac{\chi}{6\,\pi}\,\int_0^r \big\{9\,\rho(x)-p_r(x)-2p_t(x)\big\}\,x^2\,dx}_{m_{frt}}\nonumber\\+\underbrace{4\pi\,\int_0^r{\rho(x)\,x^2\,dx}}_{m_{GR}} + \underbrace{4\pi\,\beta\, \int_0^r{\theta^0_0(x)\,x^2\,dx}}_{m_{CGD}}.
\end{eqnarray}

Obviously, the limit $\chi\rightarrow 0$ and $\beta\rightarrow 0$ will provide the usual mass function expression for an anisotropic compact structure in the arena of GR. From the Eq. (\ref{eq38a}) we extra contributions due to the $f(R,T)$ and CGD scenarios as $m_{frt}(r)$ and $m_{CGD}(r)$, respectively. It is to be noted at this point that the mixture $f(R,T)$ + CGD going beyond the pure GR scope and hence clearly helps to enhance the compactness, at least from the theoretical point of view. 

It can be observed from the right hand side of Eqs. (\ref{eq27}) -- (\ref{eq29}) that the first members were separated from the new piece containing $\beta$ and the decoupler functions $h(r)$ and $g(r)$. Now, the separated field equations and their corresponding conservation law can be written as
\begin{eqnarray}\label{eq31}
&&\hspace{-0.7cm} {8\,\pi}\,\big(\rho-\hat{T}^{0}_{\ 0}\big) =\bigg[\frac{1}{r^{2}}-\frac{\xi}{r^{2}}-\frac{\xi^{\prime}}{r}\bigg],~~~~ \\\label{eq32}
&&\hspace{-0.7cm} {8\,\pi}\,\big(p_r +\hat{T}^1_1\big)= \bigg[\xi\left(\frac{1}{r^{2}}+\frac{\eta^{\prime}}{r}\right)-\frac{1}{r^{2}}\bigg] ,\\\label{eq33}
&&\hspace{-0.7cm} {8\,\pi}\,\big(p_t +\hat{T}^2_2\big)= \bigg[\frac{\xi}{4}\left(2\eta^{\prime\prime}+\eta^{\prime2}+2\frac{\eta^{\prime}}{r}\right) +\frac{\xi^{\prime}}{4}\left(\eta^{\prime}+\frac{2}{r}\right)\bigg],~~~~
\end{eqnarray}

Now Bianchi identities require the following conservation equation related to the above system of Eqs. (\ref{eq31}) -- (\ref{eq33}), i.e. $\nabla^{\mu}\big( T_{\mu\nu}+\hat{T}_{\mu\nu}\big)=0$ for when $\beta=0$ and provides  
\begin{eqnarray}\label{eq36}
p^{\prime}_r+\frac{\eta^{\prime}}{2}\,(\rho+p_r)-\frac{2}{r}\,(p_t-p_r)=\frac{\chi\,(3\rho^{\prime}-p^{\prime}_r-2p^{\prime}_t)}{6\,(4\,\pi+\chi)}.~~~~ \label{TOV} 
\end{eqnarray} 

The Eq.~(\ref{eq32}) can be termed as the modified Tolman-Oppenheimer-Volkoff (TOV)~\cite{Tolman1939,Oppenheimer1939} equation in the arena of $f(R,T)$ gravity theory.  However, it is noticeable that Eq. (\ref{eq36}) converts into the hydrostatic equilibrium condition for standard GR for$\chi=0$.
However, the corresponding solution can be given from the following static spacetime as 
\begin{equation}\label{eq34}
ds^{2} = e^{\eta(r) } \, dt^{2}-\xi^{-1}(r) dr^{2} -r^{2}(d\theta ^{2} +\sin ^{2} \theta \, d\phi ^{2}).
\end{equation} 

In this scenario, the gravitational mass for anisotropic matter distribution in $f(R,T)$-gravity can be determined by the formula as 
\begin{eqnarray} \label{eq35}
 m_0(r)=\underbrace{\frac{\chi}{6\,\pi}\,\int_0^r \big\{9\,\rho(x)-p_r(x)-2p_t(x)\big\}\,x^2\,dx}_{m_{frt}}+\underbrace{4\pi\,\int_0^r{\rho(x)\,x^2\,dx}}_{m_{GR}}.
\end{eqnarray}

Here $m_0$ is the mass function in pure $f(R,T)$ scenario. Let us now look at the factor $\beta$, so that the field equations for $\theta_{\mu\nu}$ becomes
\begin{eqnarray}\label{eq37}
&&\hspace{-0.9cm}\theta^{0}_{\ 0} =
- \frac{1}{8\pi}\bigg[\frac{h}{r^{2}}+\frac{h^{\prime}}{r}\bigg],~~~~ \\\label{eq38}
&&\hspace{-0.9cm} \theta^{1}_{\ 1}= -
 \frac{h}{8\pi}\bigg[\frac{1}{r^{2}}+\frac{\eta^{\prime}}{r}\bigg]+ \frac{\xi\,g^\prime}{r},\\\label{eq39}
&&\hspace{-0.9cm} \theta^{2}_{\ 2}=-  \frac{1}{8\pi}\bigg[\frac{h}{4}\left(2\eta^{\prime\prime}+\eta^{\prime2}+2\frac{\eta^{\prime}}{r}\right)+\frac{h^{\prime}}{4}\left(\eta^{\prime}+\frac{2}{r}\right)+\Psi(r)\bigg],~~~~
\end{eqnarray}
with
\begin{eqnarray}
\Psi(r)=\frac{\mu ^\prime h^\prime}{4}+ \frac{\mu }{4}\Big(2h^{\prime \prime}~+\beta h^{\prime 2}~+ \frac{2h^\prime}{r}~+2 \xi ^\prime h^\prime \Big). \nonumber
\end{eqnarray}

Hence, in view of Eq. (\ref{eq16}), we obtain $\nabla^{\mu}\theta_{\mu\nu}=0$ in the following conservation equation for system of Eqs. (\ref{eq37})--(\ref{eq39})
\begin{eqnarray}\label{eq40}
 \left(\theta^{1}_{\ 1}\right)^{\prime}-\frac{y^{\prime}}{2}\left(\theta^{0}_{\ 0}-\theta^{1}_{\ 1}\right)-\frac{2}{r}\left(\theta^{2}_{\ 2}-\theta^{1}_{\ 1}\right)=\frac{g^\prime}{2}(p_r+\rho).~~ 
\end{eqnarray}

In addition to this, now we use the pressure anisotropy condition in Eq. (\ref{eq28}) and (\ref{eq29}), i.e. $G^1_1=G^2_2$, which provides
 \begin{eqnarray}\label{eq41}
{\xi}\left(\frac{\eta^{\prime\prime}}{2}+\frac{\eta^{\prime2}}{4}-\frac{\eta^{\prime}}{2\,r}-\frac{1}{r^2}\right)+\frac{\xi^{\prime}\eta^{\prime}}{4}+\frac{2\,\xi^{\prime}}{r}-\frac{1}{r^2}=\hat{\Delta}, 
\end{eqnarray}
where $\hat{\Delta}={(8\pi+2\chi)}\,(p_t-p_r)$. 

One can observe here that condition (\ref{eq41}) is same as the anisotropic condition in GR. Since $f(R,T)$ gravity is an extended form of GR for the linear choice of $f(R,T)$, we can say that the solution of the field equations in $f(R,T)$ theory can be obtained by the known  solution of GR. The coupling parameter $\chi$ will only affect the matter variable. Using Eqs. (\ref{eq21})--(\ref{eq23}), we can obtain $\rho$, $p_r$ and $p_t$ form Eqs. (\ref{eq31})--(\ref{eq32}) with respect to the radial coordinate as
\begin{eqnarray}\label{eq42}
&&\hspace{-0.7cm} 8\pi \rho= \frac{1}{48 (\chi^2 + 6 \chi \pi + 8 \pi^2) r^2} \Big[48 \pi (1 -\xi^\prime r - \xi) +
 \chi \big\{16 \nonumber\\&& \hspace{0.3cm}+ \xi^\prime r (\eta^{\prime} r-16) + (4 \eta^{\prime} r + 2 \eta^{\prime\prime} r^2 + 
       \eta^{\prime2} r^2-16) \xi\big\}\Big],~~~~~\\\label{eq43}
&&\hspace{-0.7cm}8\pi p_r= \frac{1}{48 (\chi^2 + 6 \chi \pi + 8 \pi^2) r^2} \Big[48 \pi (\xi + \eta^{\prime} r \xi-1) - 
\chi \big\{16 \nonumber\\&&\hspace{0.3cm} + \xi^\prime r (8 + \eta^{\prime} r) - (16 +20 \eta^{\prime} r -2 \eta^{\prime\prime} r^2 -  \eta^{\prime2} r^2) \xi\big\}\Big],~~\\\label{eq44}
 &&\hspace{-0.7cm}8\pi p_t= \frac{1}{48 (\chi^2 + 6 \chi \pi + 8 \pi^2) r^2} \Big[ 12 \pi r (\xi^\prime (2 + \eta^{\prime} r) + (2 \eta^{\prime} \nonumber\\&& \hspace{0.3cm} + 2 \eta^{\prime\prime} r  + \eta^{\prime2} r) \xi) +  \chi \big\{8 + \xi^\prime r (4 + 5 \eta^{\prime} r) + (-8 + 8 \eta^{\prime} r\nonumber\\&&\hspace{0.3cm}  + 10 \eta^{\prime\prime} r^2 + 5 \eta^{\prime2} r^2) \xi\big\} \Big]. \end{eqnarray}

To find the contribution of $f(R,T)$, it is required to separate out  $\rho$, $p_r$ and $p_t$, because the term $\hat{T}$ in Eqs. (\ref{eq31})--(\ref{eq33}) is appearing with $\rho$, $p_r$ and $p_t$ respectively in Eqs. (\ref{eq31})--(\ref{eq33}). If we interpret the term $\rho+\hat{T}^{0}_{\ 0}$ in Eq. (\ref{eq31}) as effective seed density (similarly $p_r-\hat{T}^{1}_{1}$ in (\ref{eq32})  and $p_r-\hat{T}^{2}_{2} $ in \ref{eq33} as the effective seed pressures), then the system of the equations can be treated as the set of Einstein's field equations. The $f(R,T)$ contribution is hidden within the redefined thermodynamic quantities as mentioned in Eqs. (\ref{eq42})--(\ref{eq44}). We would like to mention here that no substantial effect has been noticed in Eqs. (\ref{eq42})--(\ref{eq44}) $f(R,T)$ since $\theta$-sector is separated by applying the CGD. Hence, in order to close the problem at least mathematically and to check the physical viability, we need to find a solution for the system $\{\theta_{\mu\nu},h,g\}$. 

Let us now define the following new physical parameters:
\begin{eqnarray}\label{eq45}
&&\hspace{0.3cm} \rho^{\text{eff}}= \rho+\beta\,\theta^0_0,\\ \label{eq46}
&&\hspace{0.3cm} p^{\text{eff}}_{r}=p_r-\beta\,\theta^1_1,\\ \label{eq47}
&&\hspace{0.3cm} p^{\text{eff}}_{t}=p_t-\beta\,\theta^2_2,
\end{eqnarray}
where the effective thermodynamic variables characterizing the matter distribution of the model. 

Here we can observe the effects of CGD on the mass function $m(r)$ from $m_{CGD}(r)$. Moreover, the effective anisotropy factor $\Delta$ can be given as  
\begin{equation}\label{eq48}
\Delta^{\text{eff}}~=~ \underbrace{(p_t-p_r)}_{\Delta_{FRT}}~ +~ \underbrace{\beta\,\, (\theta^{1}_{1}- \theta^{2}_{ 2})}_{\Delta_{CGD}}.
\end{equation}

Here we would like to highlight an important point that the CGD induced an extra contribution $\Delta_{CGD}$  in the seed anisotropy $\Delta_{FRT}$ which may enhance the anisotropy within the matter distribution. This contribution may also improve the equilibrium mechanism of the stellar system via the anisotropic force.

\subsection{Embedding Class I solution in $f(R,T)$ gravity for the seed system}\label{5.3aa}
A literature survey shows that the temporal and radial components of the chosen seed spacetime corresponds to the well-known Adler~\cite{Adler} and Finch-Skea~\cite{Finch}, respectively. Furthermore, this seed Class I spacetime was previously discussed in $f(R,T)$ gravity in the context of MGD approach~\cite{Maurya2020}. Therefore, hybridization which we are considering here is reasonable. Now, we consider the seed spacetime in $f(R,T)$ gravity, which satisfies the Karmarker condition (\ref{eq56}), is given by 
\begin{eqnarray}
 \label{eq57}
ds^{2} = (X+ B\,r^2)^2 \, dt^{2} -r^{2}(d\theta ^{2} +\sin ^{2} \theta \, d\phi ^{2}) -(1+Y r^2)\, dr^{2},~~~
\end{eqnarray}
where~$X$ and $B$ are constant and $Y=16\,A\,B^2$. 

So, the density and the pressures in $f(R,T)$ scenario [using Eqs. (\ref{eq42}) -- (\ref{eq44})] characterizing this model can be given as 
\begingroup
\small
\begin{eqnarray}\label{eq58}
&&\hspace{-0.7 cm} 8\pi\rho=\frac{2 (\chi + 3 \pi) Y (3 + Yr^2) +
 C [6 \pi Yr^2 (3 + Yr^2) + \psi_1(r)]}{6 (\chi^2 + 6 \chi \pi + 8 \pi^2) (1 + C r^2) (1 + Y r^2)^2},~~~~\\\label{eq59}
&&\hspace{-0.7 cm} 8\pi p_r=\frac{-2 Y \,\psi_3 (r) + 
 C [\chi (9 + 10 Yr^2 - 2 Y^2r^4) + \psi_2(r)]}{6 (\chi^2 + 6 \chi \pi + 8 \pi^2) (1 + C r^2) (1 + Y r^2)^2} ,\\\label{eq60}
 &&\hspace{-0.7 cm} 8\pi p_t= \frac{Y (\chi Y r^2 -6 \pi) + 
 C [6 \pi (4 + Y r^2) + \chi \psi_4(4)]}{6 (\chi^2 + 6 \chi \pi + 8 \pi^2) (1 + C r^2) (1 + Y r^2)^2},~~~
\end{eqnarray}
\endgroup
where we considered here $C=B/X$ for simplicity \\
$\psi_1(r)=\chi (3 + 8 Yr^2 + 2 r^4 Y^2)$,~~~$\psi_2(r)=6 \pi (4 + 3 Yr^2 - r^4 Y^2)$,\\
$\psi_3(r)=[\chi\,Y r^2 + 3 \pi (1 + Yr^2)]$,~~~$\psi_4(r)=(9 + 4Yr^2  +Y^2 r^4 )$. \\

Now we have already specified the $\xi$ and $\eta$, so only we need to determine the deformation functions $h(r)$ and $g(r)$ to close the $\theta$-sector completely. Therefore, we will apply two different approach to find deformation functions $h(r)$ and $g(r)$ as discussed below:

\subsubsection{Mimic constraints for the density approach for determining the deformation function $h(r)$}
In this case we will consider the mimic constraints for density 
\begin{eqnarray}\label{eq61}
\theta^0_0(r)=\rho(r),
\end{eqnarray}
which leads
\begin{eqnarray}
h^{\prime}+\frac{h}{r} =-8\pi\,r\,\rho~~\Longrightarrow~~h=-\frac{8\pi}{r}\,\int{\rho\,r^2\,dr}+\frac{F}{r}, \label{eq62}
\end{eqnarray}
where $F$ is a constant of integration. 

Now, using the Eqs. (\ref{eq58}) and (\ref{eq62}), we find the deformation function $h(r)$
\begin{eqnarray}\label{eq63}
h(r)=\frac{h_1(r)+h_2(r)\,  \tan^{-1}[\sqrt{C}\, r] - h_3(r)\,\tan^{-1}[ \sqrt{Y}\,r]}{12 (\chi^2 + 6 \chi \pi + 8 \pi^2)\,r\,(C - Y)^2 \sqrt{Y} (1 + Yr^2)}.~~
\end{eqnarray}

Here, we take the arbitrary constant $F=0$ to avoid the singularity at $r=0$. The coefficient used in $h(r)$ can be provided as
\begin{eqnarray}
&&\hspace{-0.6cm} h_1(r)= -r\, (C - Y) \,\sqrt{Y} \,[-4 (\chi + 3 \pi) Y^2 r^2  + C\, (\chi  + 4 \chi Yr^2  + 12 \pi Yr^2)],\nonumber \\
&&\hspace{-0.6cm} h_2(r)=2 \sqrt{C}\, \chi\, (3 C - 2 Y)\, \sqrt{Y}\,(1 +Y r^2),\nonumber\\
&&\hspace{-0.6cm} h_3(r)=C \chi (5 C - 3 Y) (1 +Y r^2). \nonumber
\end{eqnarray}

We would like to mention that all the calculations have been done by taking the following assumption to avoid the singularity in the expression of $f(r)$ as: (i) by expanding the $\tan^{-1}(x)$, where $x=\sqrt{C}\,r~~or~~\sqrt{Y}\,r $, upto linear term using Taylor series expansion around $x = 0$ and (ii) the integration constant involve in the solution is taken to be zero. While the other deformation function  $g(r)$ will be obtained by taking a linear equation of state (EOS) between the $\theta$-components which we discuss in the next section.

\subsubsection{Equation of state approach for determining $g(r)$}
We consider the following linear EOS in $\theta$-components to determine the function $h(r)$ as
\begin{eqnarray} \label{eq64}
&&~~ \theta^1_1 = \alpha\,\theta^0_0+\gamma,
\end{eqnarray}
where $\alpha$ and $\gamma$ are the constants. 

This EOS will provide the first order linear differential equation whose solution yields the expression for $h(r)$ as
\begin{eqnarray}
g(r)=\frac{-4\,Y\, g_1(r)\,\ln(1 + C r^2)+C [\,g_ 2 (r) + g_ 3 (r)\,]}{24\,C\, (\chi^2 + 6 \chi \pi + 8 \pi^2)\,(C-Y)},~~~~\label{eq65}
\end{eqnarray}
where
\begin{eqnarray}
&&\hspace{-0.2cm} g_1(r)=3 C (\chi + 4 \pi) - 4 (\chi + 3 \pi) Y,\nonumber\\
&&\hspace{-0.2cm}g_2(r)=r^2 \big[3 C (\chi + 4 \pi) (-(-5 + \alpha) Y + 2 \chi \gamma (2 +Y r^2 )\nonumber\\&&\hspace{1.0cm} +    4 \gamma \pi (2 + Yr^2)) -   2 Y \big\{\,3 \chi^2 \gamma (2 + Yr^2 ) \nonumber\\&& \hspace{1.0cm}+ 6 \pi ((5 - \alpha) Y + 4 \gamma \pi (2 + Yr^2)) + 
      2 \chi ((5 - \alpha) Y \nonumber\\&&\hspace{1.0cm} + 9 \gamma \pi (2 + Yr^2))\,\big\}\big]\nonumber\\
&&\hspace{-0.2cm} g_3(r)=2 \alpha (-3 C (\chi + 4 \pi) + 4 (\chi + 3 \pi) Y) \ln(1 +Y r^2).\nonumber     
\end{eqnarray}

Now the $\theta$ components can be given as
\begin{eqnarray}
&&\hspace{-0.5cm}  8\pi \theta^0_0 = \frac{2 (\chi + 3 \pi) Y (3 + Yr^2) + 
 C [6 \pi Yr^2 (3 + Yr^2) + \psi_1(r)]}{6 (\chi^2 + 6 \chi \pi + 8 \pi^2) (1 + C r^2) (1 + Y r^2)^2},~~~~~\label{eq66}\\
&&\hspace{-0.5 cm} 8 \pi \theta^1_1 = \frac{-3 C (\chi + 4 \pi) [\theta_{11}(r) - \alpha Y (3 + r^2 Y)]+\theta_{12}(r)}{12\, (\chi^2 + 6\,\chi\, \pi + 8 \pi^2) (C - Y) (1 + Y r^2)^2},\label{eq67}\\
&&\hspace{-0.5 cm} 8\pi \theta^2_2=\frac{\theta_ {21}(r) + \Psi_ {22}(r) }{2304 (\chi^2 + 6 \chi \pi + 8 \pi^2)^2 (C - Y)^2 (1 + r^2 Y)^2}.\label{eq68}~~~~
\end{eqnarray}

The coefficients used in the expressions (\ref{eq66})--(\ref{eq68}) have been mentioned in the Appendix. 

Then the deformed spacetime can be given by
\begin{eqnarray}
\label{eq69}
ds^{2} =(X+Br^2)^2\,e^{\beta\,g(r)}\, dt^{2} -r^{2}(d\theta ^{2} +\sin ^{2} \theta \, d\phi ^{2})-\frac{(1+Y r^2)}{1+\beta\,h(r)(1+Y r^2)}\, dr^{2},
\end{eqnarray}
where $h(r)$ and $g(r)$ is given in Eqs. (\ref{eq63}) and (\ref{eq65}). 

Thus, the effective quantities are given as
\begin{eqnarray}
&&\hspace{0.3cm} \rho^{\textrm{eff}}(r)=\rho(r)+\beta\,\theta^{0}_{\ 0}= (1+\beta)\,\rho(r),\\
&&\hspace{0.3cm} p_r^{\textrm{eff}}(r)=p_r(r)-\beta\,\theta^{1}_{\ 1},\label{eq72b}\\
&&\hspace{0.3cm} p_t^{\textrm{eff}}(r)=p_t(r)-\beta\,\theta^{2}_{\ 2}.
\end{eqnarray}

\section{Matching condition for the astrophysical system}
In the astrophysical system, the matching of space time geometries leads to its physical viability. Under the purview of general relativity, any spherically symmetric stellar object provides strict limiting case. It is must that the stellar distribution at the surface of the star ($r=R$) between the interior ($r<R$) and exterior ($r>R$) solution be smooth and continuous. The $f(R,T)$ theory of gravity can describe a non-minimal coupling between the geometry and the matter. The non-trivial coupling between the gravity and matter sectors can contribute the matter content of the outer space time that surrounds the compact structure. Also, the junction conditions in $f(R,T)$ gravity needs to be established. To establish this, the field equations of $f(R,T)$ gravity (\ref{eq14}) for the functional $f(R,T)=R+2\chi T$ required to be expressed in terms of the Ricci tensor and matter distribution. Also to analyze the contribution of the extended gravity on the outer space time matter distribution, we shall ignore $\theta_{\mu\nu}$ term. Now, we can establish a relation between the trace and Ricci scalar from (\ref{eq14}) as
\begin{equation}
R=-\left(8\pi+6\chi\right)T-8\chi\mathcal{P}.  \label{eq70}
\end{equation}

Subsequently, we can obtain the field equations as
\begin{equation}\label{eq71}
R_{\mu\nu}=2\left(\chi+4\pi\right)T_{\mu\nu}-2\left(2\pi+\chi\right)T\, g_{\mu\nu}-2\chi\, \mathcal{P}\, g_{\mu\nu}.  
\end{equation}

Now, to get an idea on the matching condition, it is informative to know the contribution from $T_{\mu\nu}$ and $f(R,T)$ gravity sector inside the compact object. For brevity, we consider $T_{\mu\nu}=0$ so that the trace, energy density and the pressures terms vanish. So, without loss of generality, we consider the geometry of outer space time as the Schwarzschild exterior metric. This is possible because of the linear relation between the Ricci scalar and trace of the energy momentum tensor(\ref{eq13}). If the gravity-matter coupling were represented by the functional $f(R,T)$ such that non-linear terms would exist between them and at a result obtaining $R$ and $T$ explicitly become difficult. The minimal coupling further allows the decoupling of $\theta_{\mu\nu}$ and seed source. would violate. Hence, the Schwarzschild metric as the exterior space time that  surrounds the collapsed configuration. As usual this solution is given by
\begin{eqnarray}
ds^2_+= - \bigg(1-\frac{2{{M}}}{r}\bigg)^{-1} dr^2-r^2(d\theta^2-\sin^2\theta\,d\phi^2)+\bigg(1-\frac{2{{M}}}{r}\bigg)\,dt^2.~\label{eq72}
\end{eqnarray} 

In this connection, the deformed solution can be given by the most general interior spacetime as
\begin{eqnarray}\label{eq73}
ds^{2}_{-}=-\left[\xi(r)+\beta h(r)\right]^{-1}dr^{2}-r^2(d\theta^2-\sin^2\theta\,d\phi^2)+ e^{\eta(r)+\beta g(r)}dt^{2}. 
\end{eqnarray}

Now, at the boundary $\Sigma$, the matching of the geometries can be performed smoothly between the the outer manifold $ds^2_+$ (\ref{eq72}) and the inner manifold $ds^2_-$ (\ref{eq73}) as per the junction conditions. Joining both the geometries at the boundary is the continuity equation. This will provide the first and second fundamental forms across the surface $\Sigma$. In the first fundamental form, the inner geometry described by the metric tensor $g_{\mu\nu}$ induced by $ds^2_-$ and $ds^2_+$ on the interface, which can be written as
\begin{eqnarray}
&&g^{-}_{00}|_{r=r_b}=g^{+}_{00}|_{r=r_b}~~ \text{and}~~  g^{-}_{33}|_{r=r_b}=g^{+}_{33}|_{r=r_b}. \label{eq74}
\end{eqnarray}

Explicitly it reads
\begin{eqnarray}
 \xi(r_b)+\alpha\,h(r_b)&=&\bigg(1-{\frac {2{{M}}}{r_b}}\bigg),~\label{eq75}\\
e^{\eta(r_b)+\beta g(r_b)} &=&\bigg(1-{\frac {2{{M}}}{r_b}}\bigg),~\label{eq76}
 \end{eqnarray}
 where
 \begin{eqnarray}
 M=m(r_b)=m_0(r_b)+4\pi\,\beta\, \int_0^r{\theta^0_0(x)\,x^2\,dx}.
 \end{eqnarray}
 
Furthermore, the second fundamental form is associated with the continuity of the extrinsic curvature $K_{\mu\nu}$ through the $\mathcal{M}^{-}$ and $\mathcal{M}^{+}$ on $\Sigma$. Then by matching of  interior ($\mathcal{M}^{-}$) and outer ($\mathcal{M}^{+}$) manifold at the $\Sigma$ gives
\begin{equation}
\left[p^{(\text{eff})}_{r}(r)\right]_{\Sigma}=\left[p(r)-\alpha\theta^{1}_{1}(r)\right]_{\Sigma}=0.\label{eq77}
\end{equation}

Here, it is noted that the $\theta$-sector could in principle introduce some modifications on the outer spacetime and matter content. Due to above fact, the second fundamental form (\ref{eq77}) can be written in the following from as 
  {\begin{equation}\label{eq78}
  \begin{split}
 p(r_b)-\alpha\,[\theta^{1}_{\ 1}(r_b]^{-} =   -\alpha\,[\theta^{1}_{\ 1}(r_b)]^{+},~
 \end{split}
  \end{equation}}
where $p_r(r_b)=p_r^{-}(r_b)$. 

After substituting the value of $(\theta^{1}_{\ 1})^{-}(r_b)$ from Eq. (\ref{eq38}), the above expression reads
\begin{eqnarray}
{p}_r(r_b)+\beta\,\bigg[\frac{h}{8\pi}\,\bigg(\frac{y^{\prime}}{r}+\frac{1}{r^2}\bigg)+\frac{\mu\,h^{\prime}}{r}\bigg]_{r=r_b} =-\beta\,(\theta^1_1)^{+}(r_b),~~~\label{eq79}
\end{eqnarray}
where $y^{\prime}\equiv \partial_r\,y^{-}$. 

In order to find out $(\theta^1_1)^{+}(r_b)$ in (\ref{eq49}), we employ Eqs.(\ref{eq38}), (\ref{eq72}) and (\ref{eq76}), obtaining
\begin{eqnarray}
{p}_r(r_b)+\beta\,\bigg[\frac{h(r_b)}{8\pi}\,\bigg(\frac{y^{\prime}(R)}{R}+\frac{1}{R^2}\bigg)+\frac{\xi(r_b)\,g^{\prime}(r_b)}{8\pi R}\bigg]\nonumber\\ =\frac{\beta\,h^{\ast}(r_b)}{8\pi}\Bigg[\frac{2M}{R^2\,\big(R-2M\big)}+\frac{1}{R^2}\Bigg]
 +\beta\,\frac{\big[g^{\ast}(r_b)\big]^{\prime}}{8\pi}\,\bigg(\frac{R-2M}{R^2}\bigg),~~~\label{eq80}
 \end{eqnarray} 
where $h^{\ast}(r_b)$ and $g^{\ast}(r_b)$ denote the deformation functions for exterior solution under the extra source $\theta_{\mu\nu}$, which can be given by following in the exterior metric as
\begin{eqnarray}
ds^2_+= - \bigg(1-\frac{2{{M}}}{r}+\beta\,g^{\ast}\bigg) dt^2+r^2(d\theta^2-\sin^2\theta\,d\phi^2)+\bigg(1-\frac{2{{M}}}{r}+\beta\,h^{\ast}\bigg)^{-1}\,dr^2.~~\label{eq85a}
\end{eqnarray} 

If this the exterior solution (\ref{eq85a}) is given by the Schwarzschild solution (\ref{eq72}), then we must put $h^{\ast}(r_b)=0$ and $g^{\ast}(r_b)=0$ in (\ref{eq85a}). Then the Eq. (\ref{eq80}) yields the following
\begin{eqnarray}
p(r_b)+\frac{h(r_b)}{8\pi}\,\bigg(\frac{\eta^{\prime}(r_b)}{R}+\frac{1}{R^2}\bigg)-\frac{ \xi (r_b) g^\prime (r_b)}{8 \pi r_b}=0.~~~\label{eq81}
 \end{eqnarray} 
 
The above condition (\ref{eq81}) can be also written as
\begin{eqnarray}
\label{eq62a}
&& p(r_b)-\alpha\,\big(\theta^{1}_1(r_b)\big)^{-}=0.
\end{eqnarray}

The constants involves in the solutions will be determined by the necessary and sufficient conditions (\ref{eq75}), (\ref{eq76}) and (\ref{eq81}). So, by using the boundary conditions we have obtained the values of the constants
\begin{eqnarray}
&&\hspace{-0.9cm} C=\frac{-2Y \big[\chi\,Y\,r_b^2  + 3 \pi (1 + r_b^2 Y)\big]+2\beta\,C_{11}(r_b)}{-6 \beta \chi^2 \gamma (r_b + r_b^3 Y)^2+C_{22}(r_b)},\\
&&\hspace{-0.9cm} M=M_0-\frac{\beta\,r}{2}\,h(r_b)=\frac{YR^3}{2(1+YR^2)}-\frac{\beta\,R\,h(R)}{2},
\end{eqnarray}
where $M_0=m_0(r_b)$. Also, we are avoiding to write the expression for the constant $B$ due to its long cumbersome form.

\section{Physical analysis of the gravitational decoupling solution for $f(R,T)$ gravity}

\subsection{Regularity conditions}
It is to note that the regular behavior of the solution depends on the deformed metric functions $e^{\lambda}$ and $e^{y}$ which are $e^{\lambda(0)}=1$ and $e^{y(0)}>0$ together with the monotonic increasing function of $r$ to describe the realistic objects. In the present case we see that $e^{\lambda(r)}$ and $e^{y}$ are solely dependent on the parameters $\xi,~\eta,~h,~g$. Since $\xi$ and $\eta$ are metric function corresponding to the seed space-time which are already well behaved, then only testing of the physical validity of the deformation functions $h(r)$ and $g(r)$ are required. At the centre $r=0$, $f(r)$ and $g(r)$ must be freed from singularity and $f(r)$ must be vanished to preserve $e^{\lambda(0)}=1$. 

In addition to the above, the following conditions also must be satisfied: 
   \subsubsection*{Case I: For $\beta>0$}  
    \begin{enumerate}
    \item if $h(r)\geq0$, $g(r)\geq0$ and for all $r\in[0,R]$, both are increasing, then the deformed metric function $e^{\lambda(r)}>0$, $e^{y(r)}>0$, mass function $m(r)>0$ are also increasing when the growth of $\xi(r)$ is faster than $h(r)$.
   \item if $h(r)\leq0$, $g(r)\leq0$ and for all $r\in[0,R]$, both are decreasing, then the deformed metric function $e^{\lambda(r)}>0$, $e^{y(r)}>0$, mass function $m(r)>0$ are increasing when the growth of $\eta(r)$ is faster than $g(r)$. 
    \item if $h(r)\le 0$, $g(r)\ge0$ for all $r\in[0,R]$ then the deformed metric function $e^{\lambda(r)}>0$, $e^{y(r)}>0$, mass function $m(r)>0$ and increasing automatically.   
    \item if $h(r)\ge 0$, $g(r)\le0$ for all $r\in[0,R]$, then the growth of $\xi(r)$ and $\eta(r)$ must be respectively faster than $h(r)$ and $g(r)$, to maintain the deformed metric function $e^{\lambda(r)}>0$, $e^{y(r)}>0$ and the mass function $m(r)>$ and its increasing behaviour.
    \end{enumerate}

%%%%%%%%%%%%%%%%%%%%%%%%%%%%%%%%%%%%%%%%%%%%%%%%%%%%%%%%%%%%%%%
\begin{figure}[tbp]
\centering
\includegraphics[width=6.5cm]{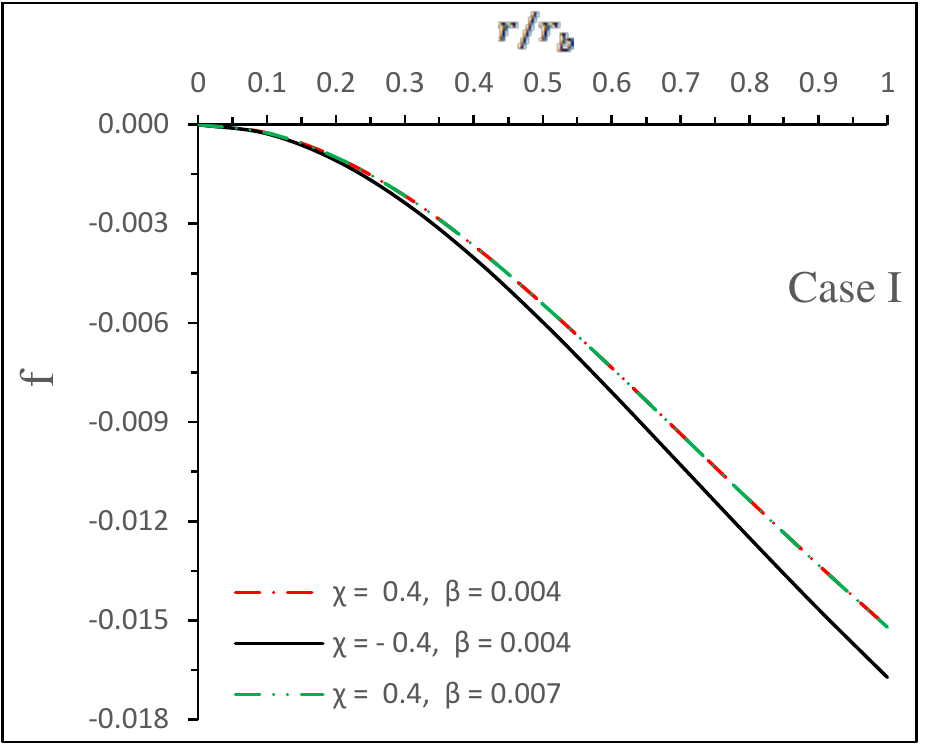}
\includegraphics[width=6.5cm]{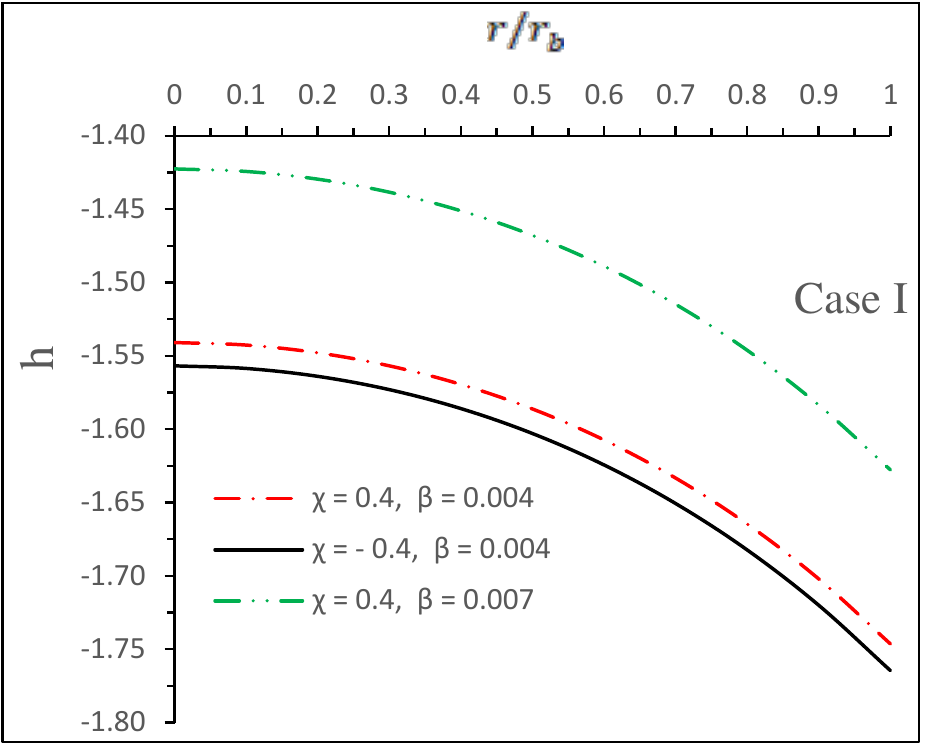} 
\caption{Variation of the radial deformation function $f(r)$ and temporal deformation function $h(r)$ with respect to the radial coordinate $r/r_b$. For plotting of this figure, we use the numerical values of the constants as $\alpha=1.4$, $\gamma=-0.002$, $\frac{M_0}{R}=0.2$, and $Y=0.005$. Henceforth we shall use this same data set for plotting other figures.}\label{f1}
\end{figure}
%%%%%%%%%%%%%%%%%%%%%%%%%%%%%%%%%%%%%%%%%%%%%%%%%%%%%%%%%%%%%%

%%%%%%%%%%%%%%%%%%%%%%%%%%%%%%%%%%%%%%%%%%%%%%%%%%%%%%%%%%%%%%
\begin{figure}[tbp]
\centering
\includegraphics[width=6.5cm]{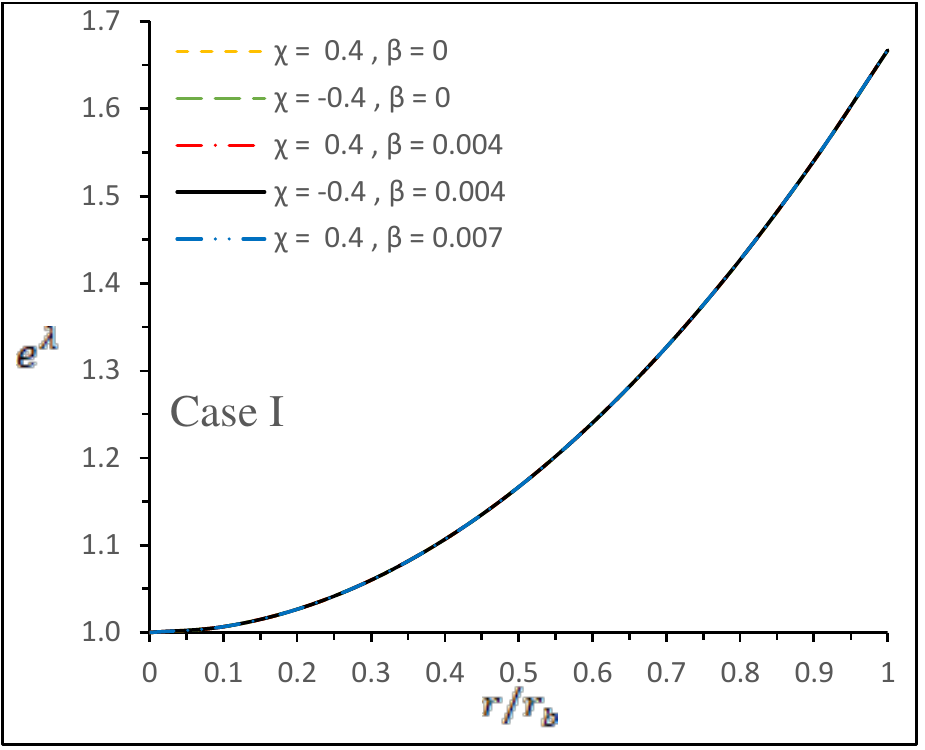}
\includegraphics[width=6.5cm]{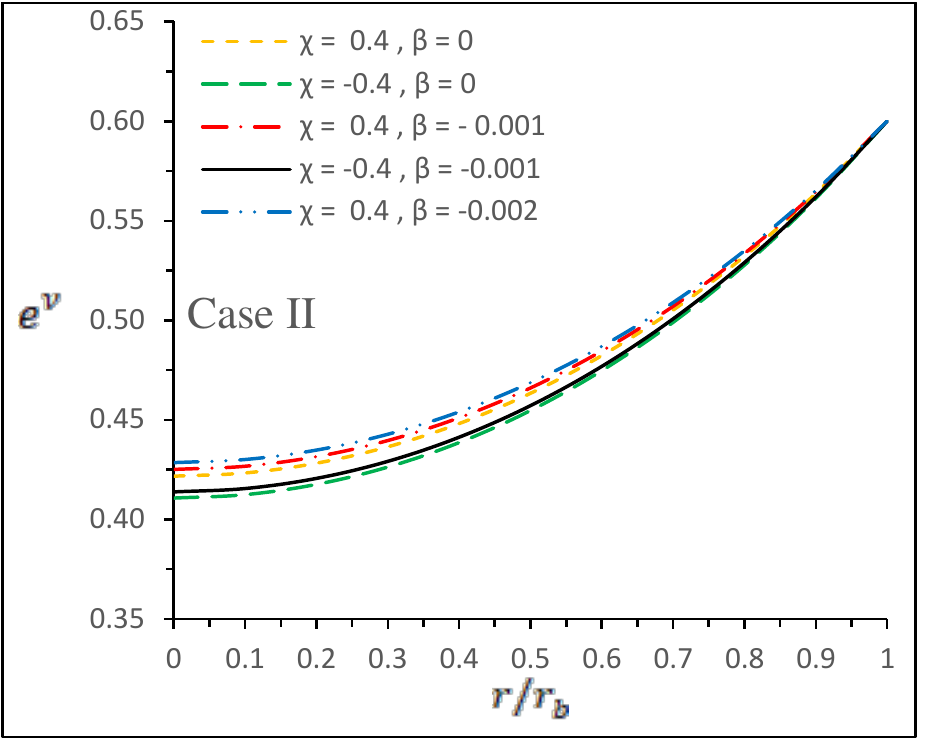} 
\caption{Variation of the metric functions $e^{\lambda}$ and $e^{\nu}$  with respect to the radial coordinate $r/r_b$.  }\label{f2}
\end{figure}
%%%%%%%%%%%%%%%%%%%%%%%%%%%%%%%%%%%%%%%%%%%%%%%%%%%%%%%%%%%%%%

%%%%%%%%%%%%%%%%%%%%%%%%%%%%%%%%%%%%%%%%%%%%%%%%%%%%%%%%%%%%%%
\begin{figure}[tbp]
\centering
\includegraphics[width=6.5cm]{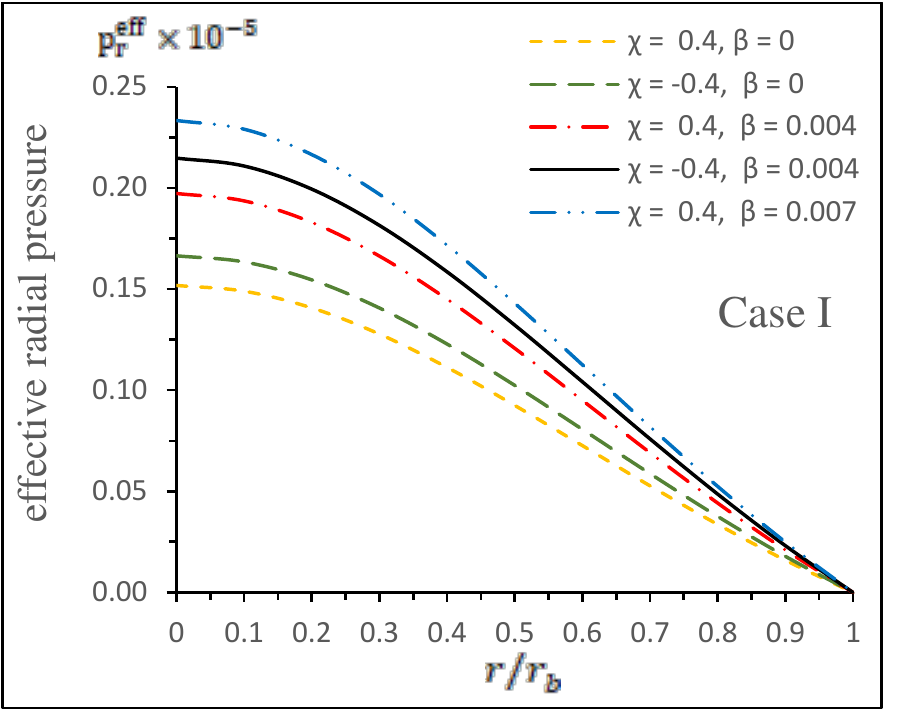}
\includegraphics[width=6.5cm]{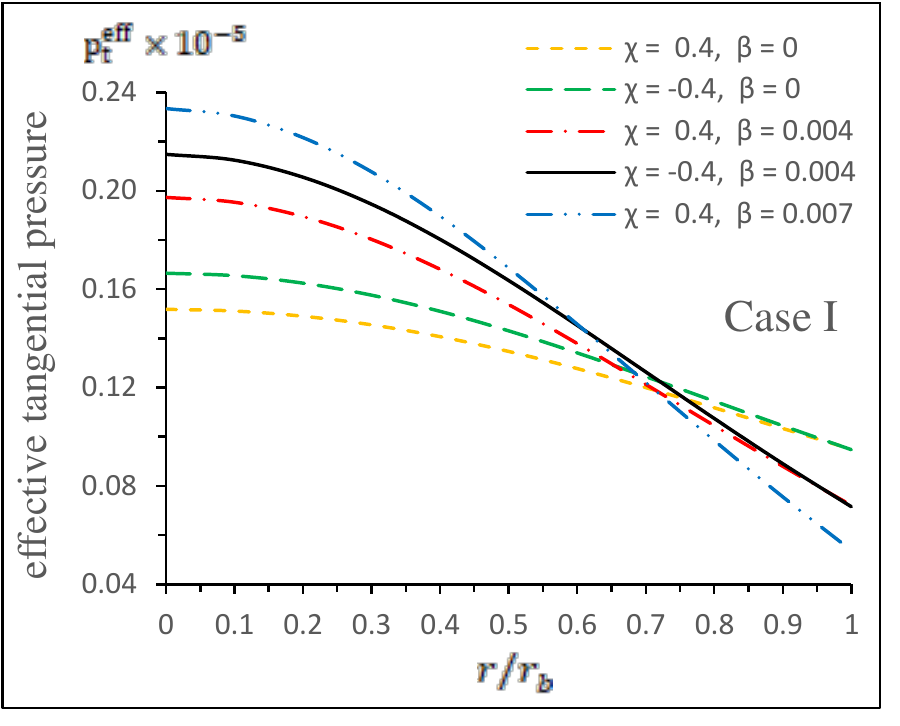} 
\caption{The left panel shows the effective radial pressure ($p^{\text{eff}}_r$) and the right panel shows the effective tangential pressure ($p^{\text{eff}}_t$) with respect to the radial coordinate $r/r_b$.}\label{f3}
\end{figure}
%%%%%%%%%%%%%%%%%%%%%%%%%%%%%%%%%%%%%%%%%%%%%%%%%%%%%%%%%%%%%%

%%%%%%%%%%%%%%%%%%%%%%%%%%%%%%%%%%%%%%%%%%%%%%%%%%%%%%%%%%%%%%
\begin{figure}[tbp]
\centering
\includegraphics[width=6.5cm]{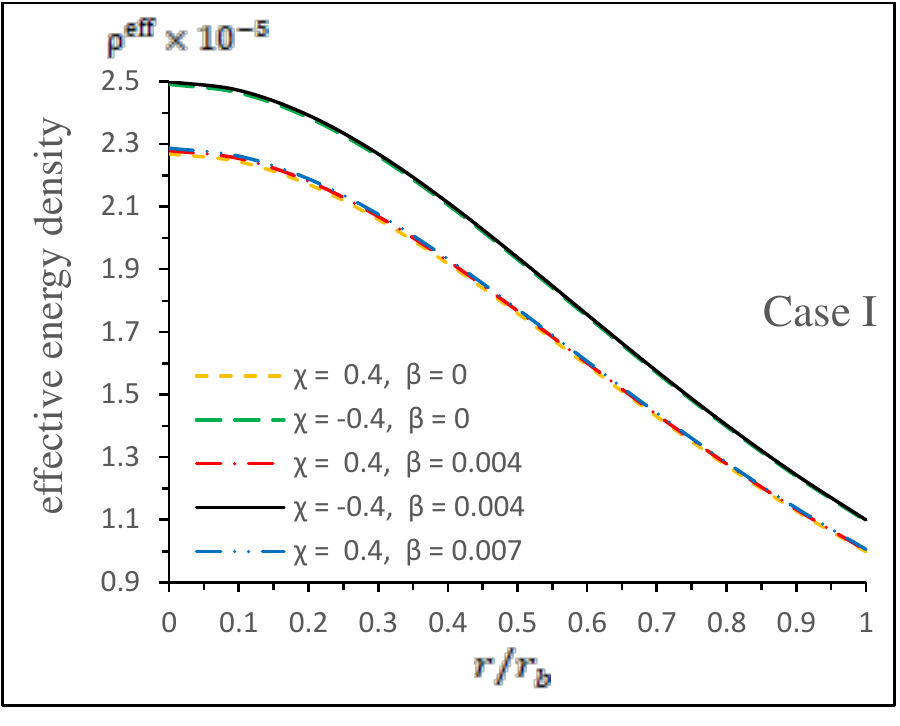}
\includegraphics[width=6.5cm]{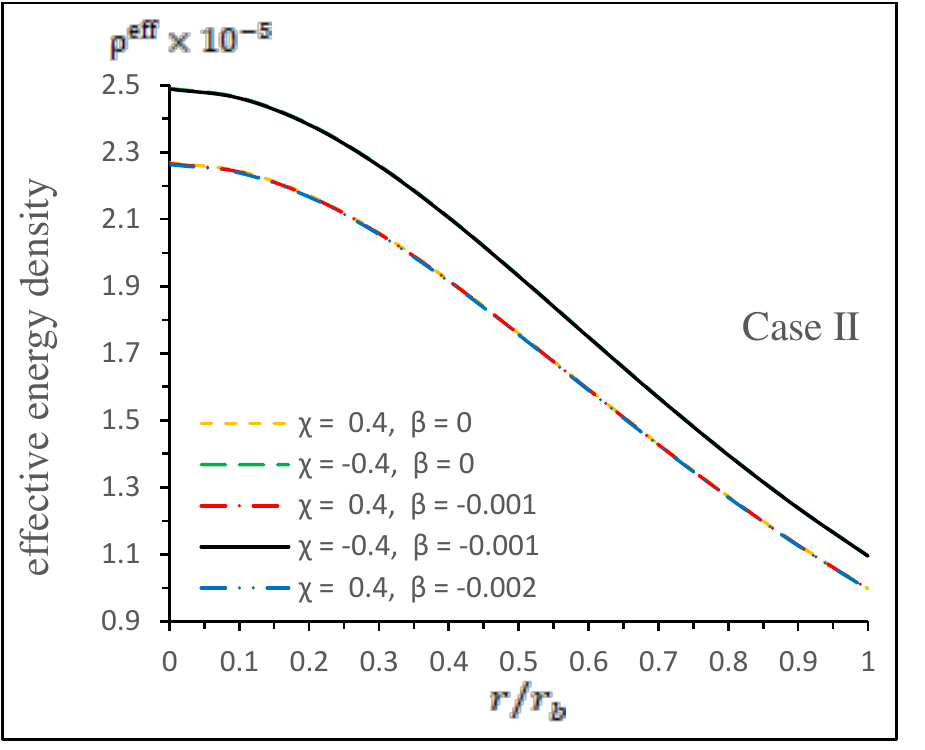}
\caption{The left panel shows the effective density $\rho^{\text{eff}}_r$ for Case I and the right panel shows the effective density $\rho^{\text{eff}}_r$  for Case II with respect to the radial coordinate $r/r_b$.}\label{f4}
\end{figure}
%%%%%%%%%%%%%%%%%%%%%%%%%%%%%%%%%%%%%%%%%%%%%%%%%%%%%%%%%%%%%%

%%%%%%%%%%%%%%%%%%%%%%%%%%%%%%%%%%%%%%%%%%%%%%%%%%%%%%%%%%%%%%
\begin{figure}[tbp]
\centering
\includegraphics[width=6.5cm]{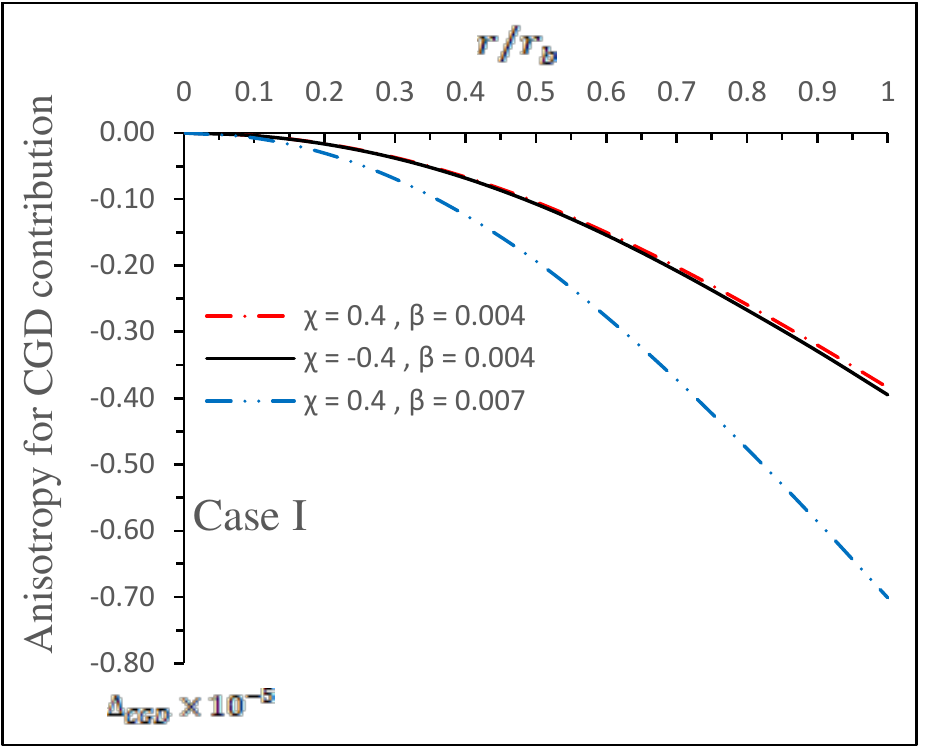}
\includegraphics[width=6.5cm]{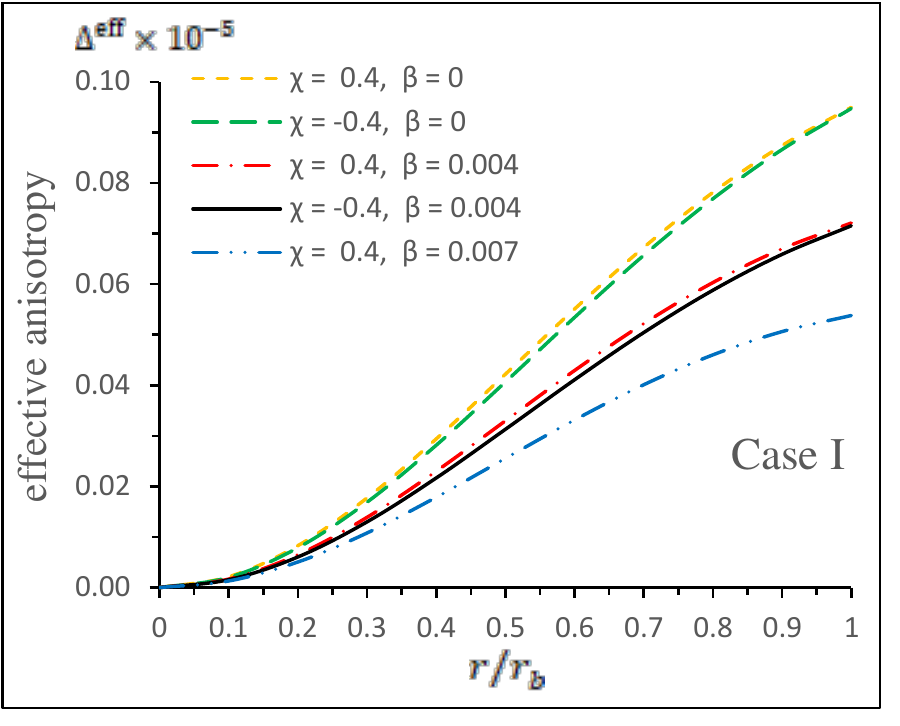}
\caption{The left panel shows the anisotropy for the CGD contribution  ($\Delta_{CGD}$) and the right panel shows the effective anisotropy ($\Delta^{\text{eff}}$) with respect to $r/r_b$. }\label{f5}
\end{figure}
%%%%%%%%%%%%%%%%%%%%%%%%%%%%%%%%%%%%%%%%%%%%%%%%%%%%%%%%%%%%%%

%%%%%%%%%%%%%%%%%%%%%%%%%%%%%%%%%%%%%%%%%%%%%%%%%%%%%%%%%%%%%%
\begin{figure}[tbp]
\centering
\includegraphics[width=6.5cm]{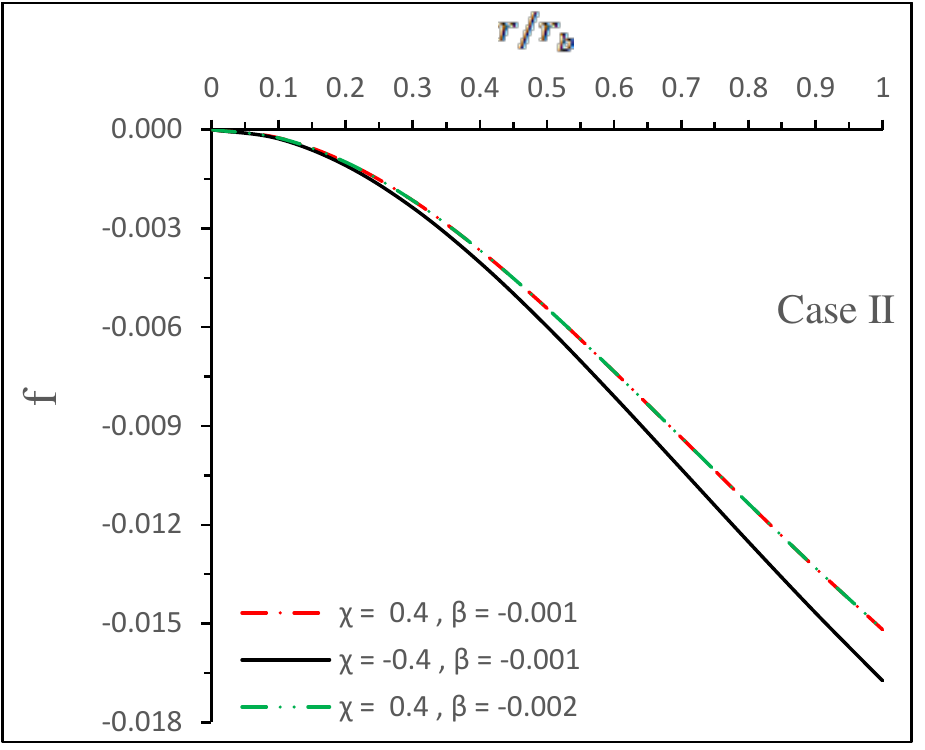}
\includegraphics[width=6.5cm]{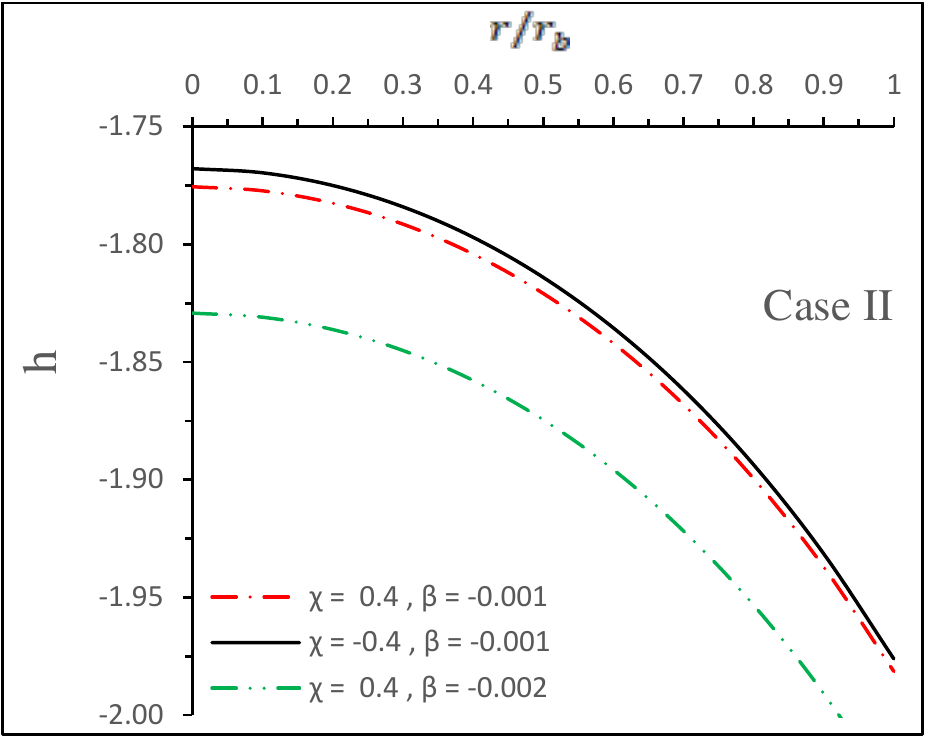} 
\caption{Variation of the radial deformation function, $f(r)$, and the temporal deformation function, $h(r)$ with respect to the radial coordinate $r/r_b$. For plotting of this figure, we use the numerical values of the constants as $\alpha=1.4$, $\gamma=-0.002$, $\frac{M_0}{R}=0.2$, and $Y=0.005$. Henceforth we shall use this same data set for plotting other figures. }\label{f6}
\end{figure}
%%%%%%%%%%%%%%%%%%%%%%%%%%%%%%%%%%%%%%%%%%%%%%%%%%%%%%%%%%%%%%

%%%%%%%%%%%%%%%%%%%%%%%%%%%%%%%%%%%%%%%%%%%%%%%%%%%%%%%%%%%%%%
\begin{figure}[tbp]
\centering
\includegraphics[width=6.5cm]{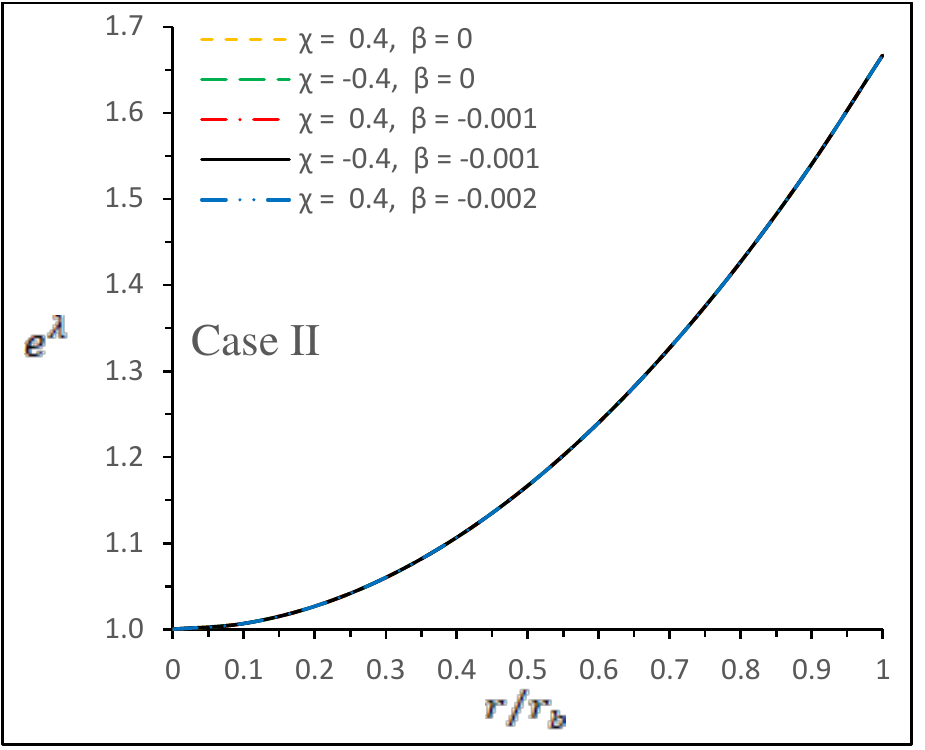}
\includegraphics[width=6.5cm]{nu2.pdf} 
\caption{Variation of the metric functions $e^{\lambda}$ and $e^{\nu}$ with respect to the radial coordinate $r/r_b$. } \label{f7}
\end{figure}
%%%%%%%%%%%%%%%%%%%%%%%%%%%%%%%%%%%%%%%%%%%%%%%%%%%%%%%%%%%%%%

%%%%%%%%%%%%%%%%%%%%%%%%%%%%%%%%%%%%%%%%%%%%%%%%%%%%%%%%%%%%%%
\begin{figure}[tbp]
\centering
\includegraphics[width=6.5cm]{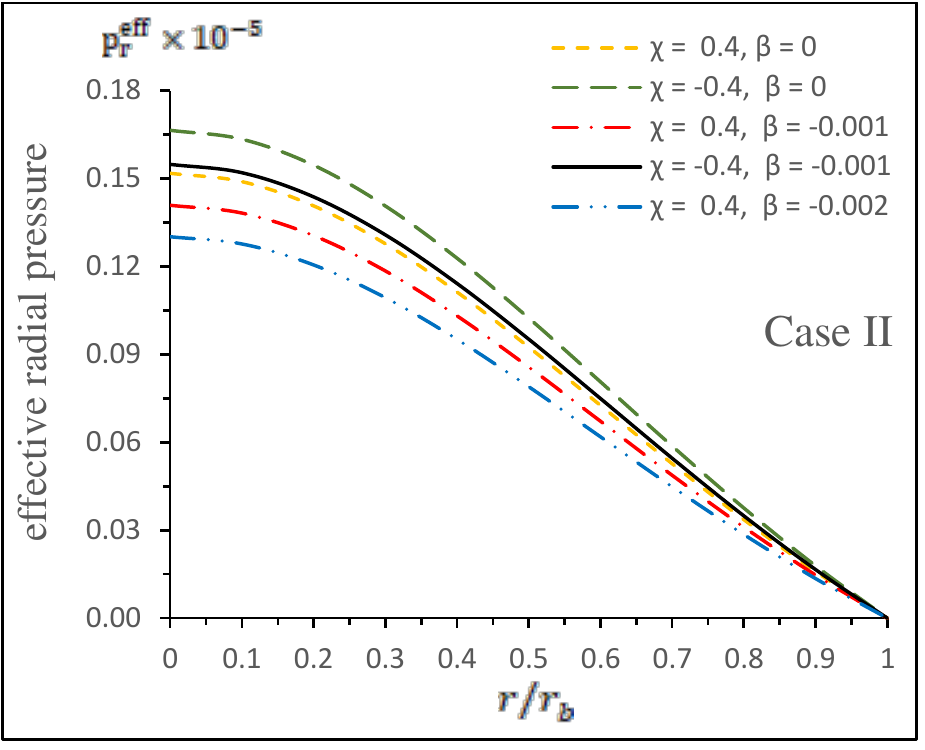}
\includegraphics[width=6.5cm]{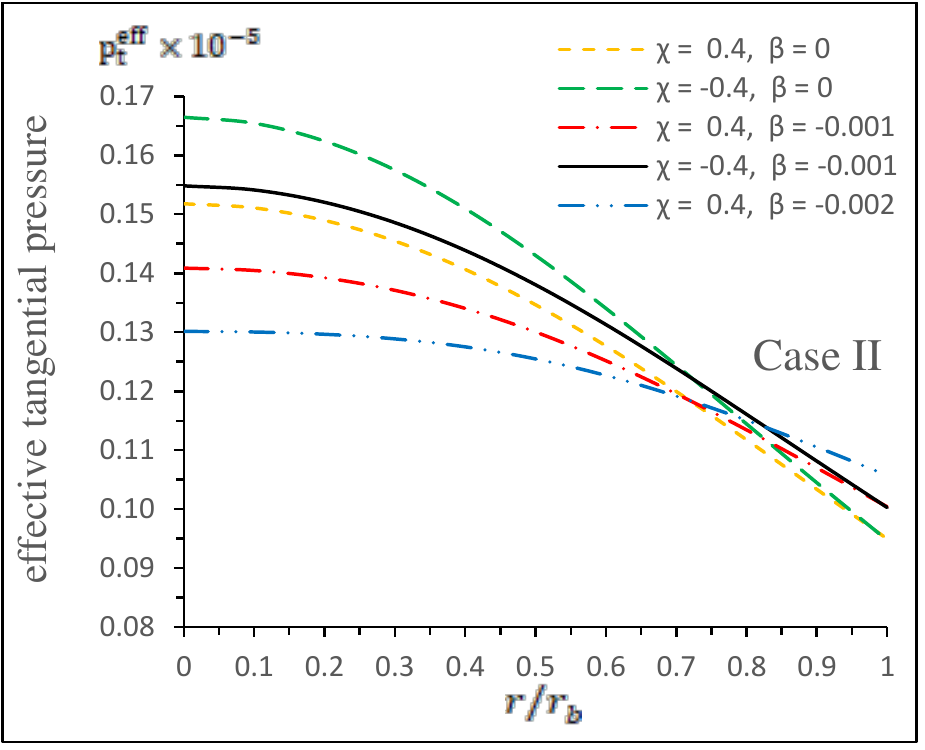} 
\caption{The left panel shows the effective radial pressure ($p^{\text{eff}}_r$) and the right panel shows the effective tangential pressure ($p^{\text{eff}}_t$) with respect to the radial coordinate $r/r_b$. }\label{f8}
\end{figure}
%%%%%%%%%%%%%%%%%%%%%%%%%%%%%%%%%%%%%%%%%%%%%%%%%%%%%%%%%%%%%%

%%%%%%%%%%%%%%%%%%%%%%%%%%%%%%%%%%%%%%%%%%%%%%%%%%%%%%%%%%%%%%
\begin{figure}[tbp]
\centering
\includegraphics[width=6.5cm]{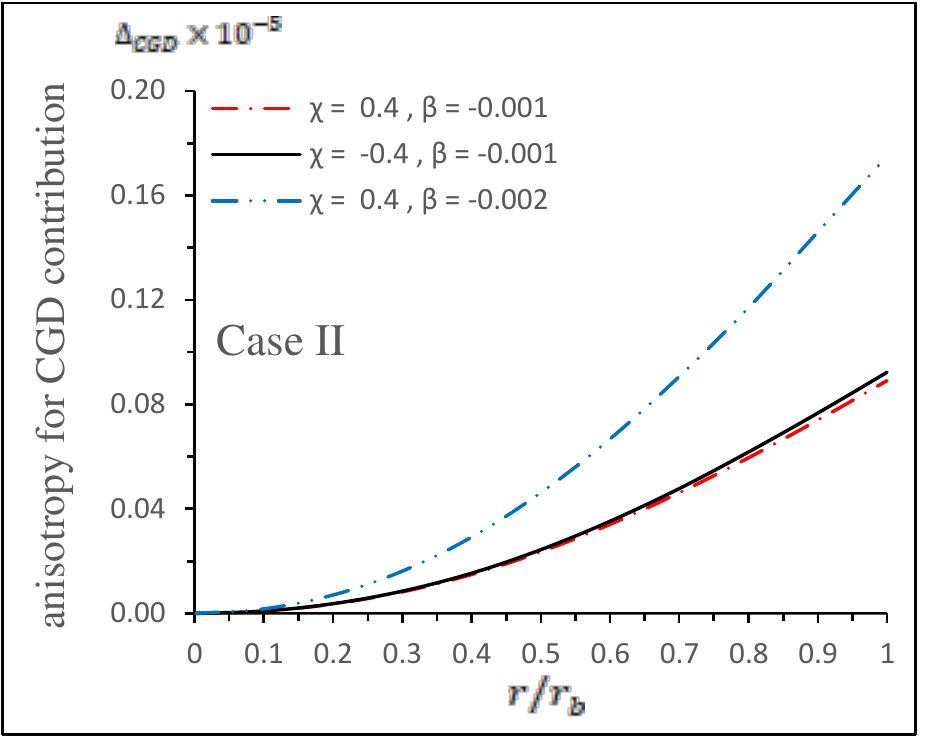}
\includegraphics[width=6.5cm]{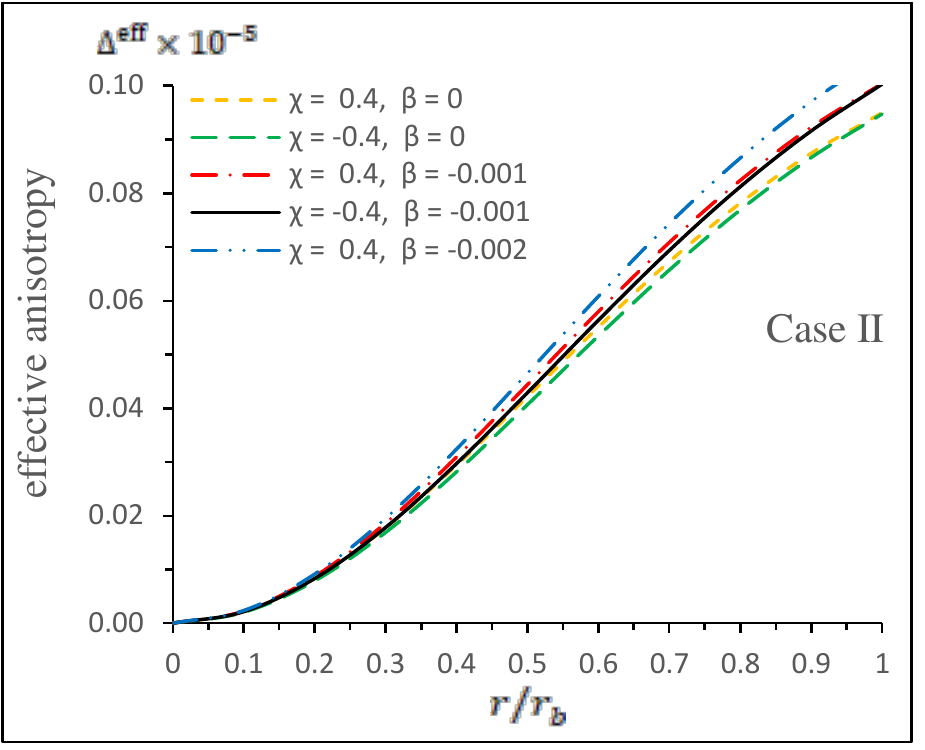}
\caption{The left panel shows the anisotropy for the CGD contribution  ($\Delta_{CGD}$) and the right panel shows the effective anisotropy ($\Delta^{\text{eff}}$) with respect to $r/r_b$. }\label{f9}
\end{figure}
%%%%%%%%%%%%%%%%%%%%%%%%%%%%%%%%%%%%%%%%%%%%%%%%%%%%%%%%%%%%%%

 \subsubsection*{Case II: For $\beta<0$}  
 \begin{enumerate}
    \item For non-negative and increasing $h(r)$ and $g(r)$, $\forall$ $r\in[0,R]$, the deformed metric function $e^{\lambda(r)}$, $e^{y(r)}$ and the mass function $m(r)$ will be positive and increasing, when the growth of $\eta(r)$ is higher than $g(r)$. 
    \item if $h(r)$ and $g(r)$ are non-positive and decreasing for all $r\in[0,R]$ then growth of $\xi(r)$ must higher than $h(r)$ in order to preserve the increasing and positive behavior of $e^{\lambda(r)}$, $e^{y(r)}$ and the mass function $m(r)$.   
  \item if $h(r)\le 0$ and $g(r)\ge0$ , $\forall$ $r\in[0,R]$ then the deformed metric function $e^{\lambda(r)}$, $e^{y(r)}$ and the mass function $m(r)$ will be positive and increasing, if the growth of $\xi(r)$ and $\eta(r)$ are higher than $h(r)$ and $g(r)$, respectively.
    \item if $h(r)\ge 0$ and $g(r)\le0$, $\forall$ $r\in[0,R]$, then it yields positive and increasing behavior of $e^{\lambda(r)}$, $e^{y(r)}$ and $m(r)$, automatically.      
\end{enumerate}

\subsection{Adiabatic Index}
The adiabatic index $\Gamma$ is defined as the ratio of two specific heats~\cite{Hillebrandt1976} as follows:
\begin{eqnarray}
\Gamma &=& \frac{\rho+p_r}{p_r}\left[\frac{dp_r}{d\rho}\right].  \label{eq-adi1}
\end{eqnarray}

This provides a tool to study the density profile as well as the stiffness of EOS of a spherically symmetric system~\cite{Harrison1965,Haensel2007,Deb2019a,Deb2019b,Biswas2020,Biswas2021}. It is argued that $\Gamma$ can play a key to explain the dynamical stability of a stellar system under the application of an infinitesimal radial adiabatic perturbation ~\cite{Chandrasekhar1964,Bardeen1966,Wald1984,Knutsen1988,Herrera1997,Horvat2011,Doneva2012,Mak2013,Silva2015,TO2021}. 
It is worthy to mention that $\Gamma>\frac{4}{3}$ is prescribed for a stable Newtonian sphere, $\Gamma=\frac{4}{3}$ gives rise to a neutral equilibrium~\cite{Bondi1964}. In the scenario of relativistic fluid distribution, some extra term will involve which may produce some correction in previous bound, which can be given as \cite{r37,r38}
\begin{equation}\label{eq85}
\Gamma<\frac{4}{3}+\left[\frac{1}{3}\kappa\frac{\rho_{0}p_{r0}}{|p^{\prime}_{r0}|}r+\frac{4}{3}\frac{\left(p_{t0}-p_{r0}\right)}{|p^{\prime}_{r0}|r}\right]_{max},    
\end{equation}
where $\rho_{0}$, $p_{r0}$ and $p_{t0}$ are called the initial density, radial and tangential pressure when the matter distribution is in static equilibrium. The first term in bracket quantity in above inequality describe the relativistic corrections to the Newtonian perfect fluid while the last term present in the bracket due to anisotropy.

%%%%%%%%%%%%%%%%%%%%%%%%%%%%%%%%%%%%%%%%%%%%%%%%%%%%%%%%%%%%%%
\begin{figure}[tbp]
\centering
\includegraphics[width=6.5cm]{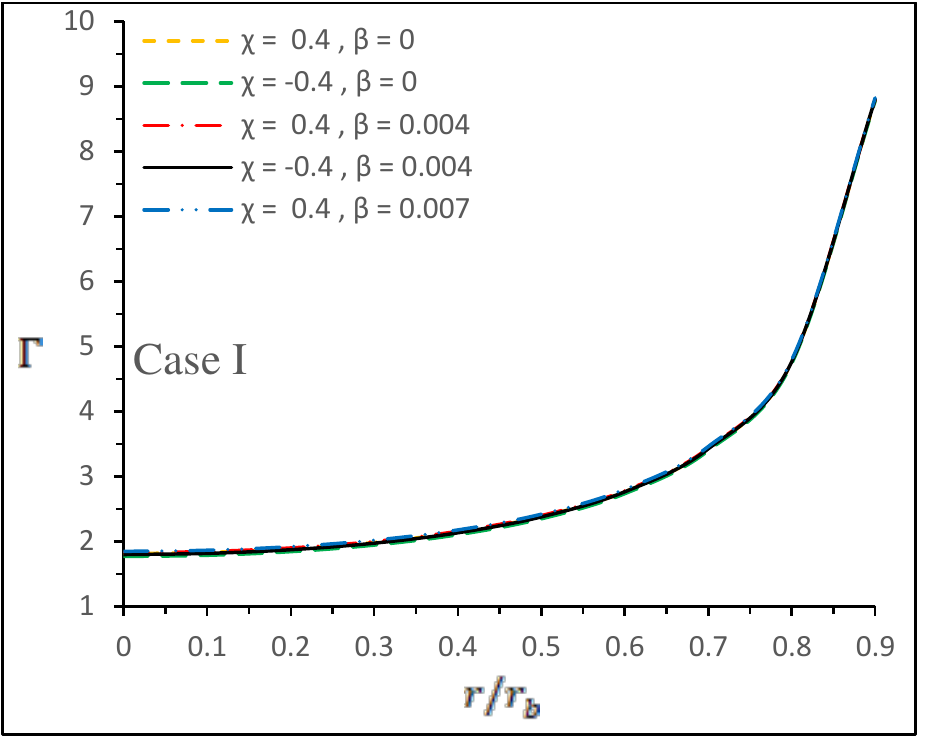}
\includegraphics[width=6.5cm]{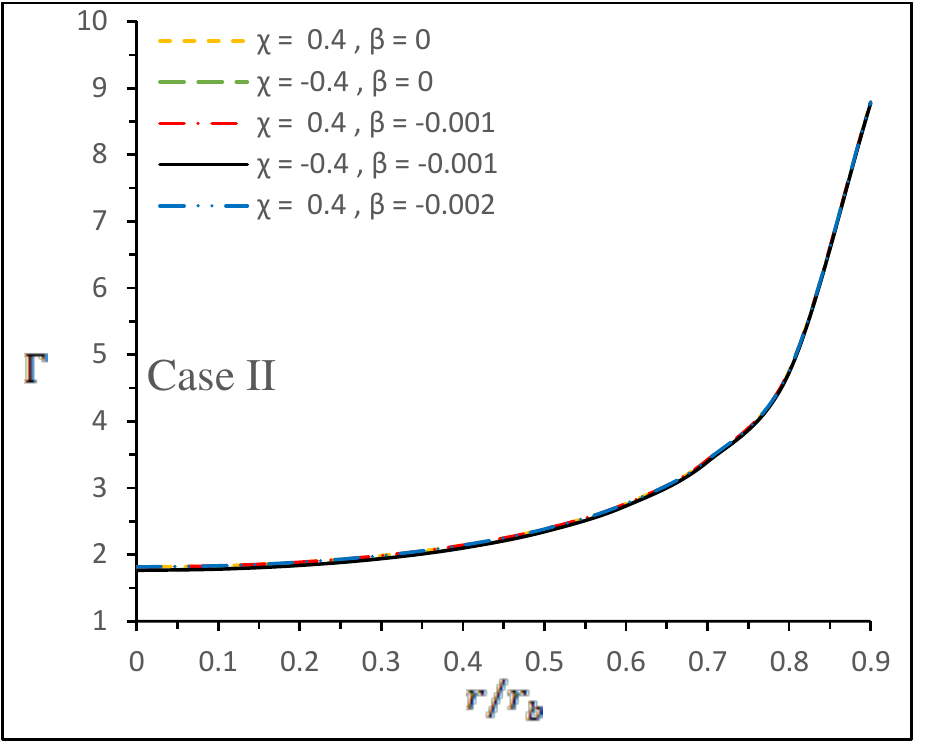}
\caption{Variation of the adiabatic index ($\Gamma$) with respect to the radial coordinate $r/r_b$. }\label{f10}
\end{figure}
%%%%%%%%%%%%%%%%%%%%%%%%%%%%%%%%%%%%%%%%%%%%%%%%%%%%%%%%%%%%%%

%%%%%%%%%%%%%%%%%%%%%%%%%%%%%%%%%%%%%%%%%%%%%%%%%%%%%%%%%%%%%%
\begin{figure}[tbp]
\centering
\includegraphics[width=6.5cm]{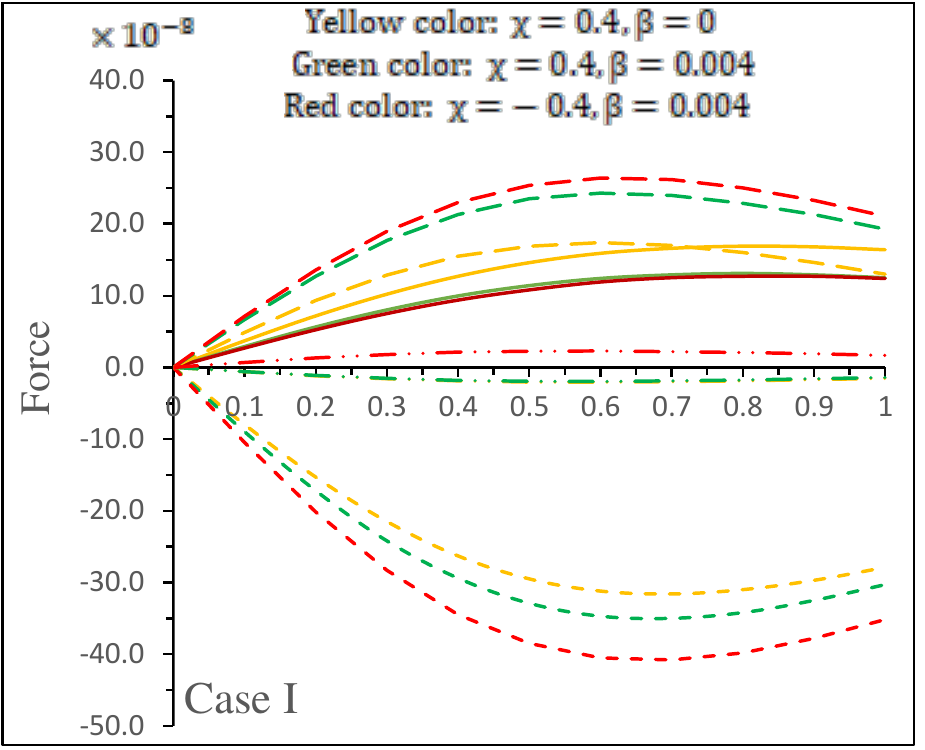}
\includegraphics[width=6.5cm]{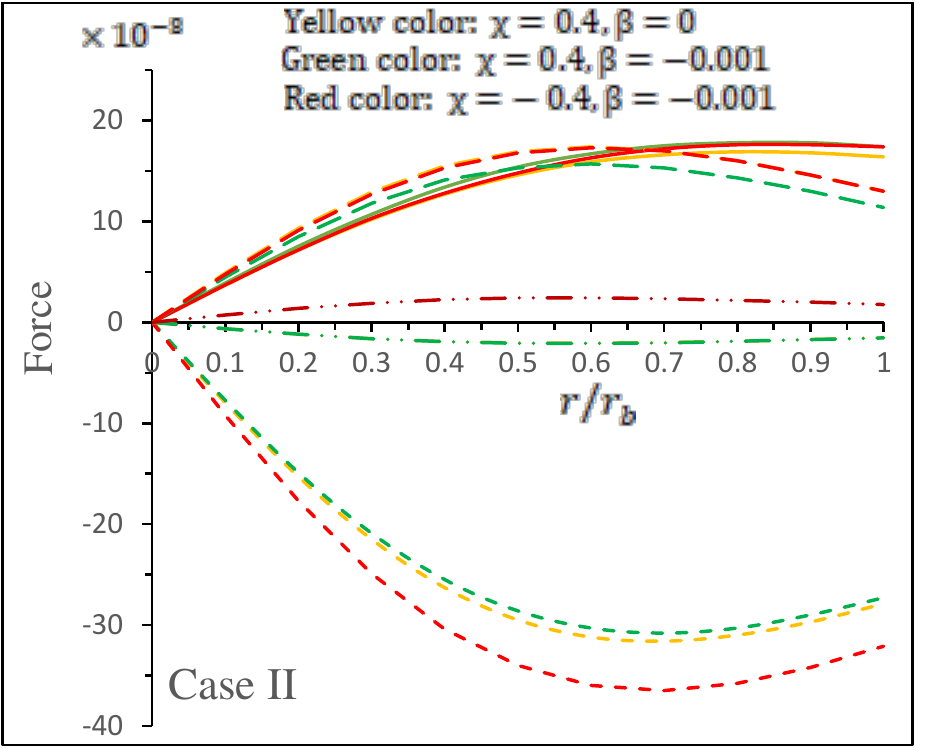}
\caption{Variation of different forces as follows: $F_g$-short dash curves, $F_h$-long dashed curves, $F_a$-Solid curves, and $F_\chi$-long dot dashed curves with respect to the radial coordinate $r/r_b$ for Case I (left panel) and Case II (right panel). }\label{f11}
\end{figure}
%%%%%%%%%%%%%%%%%%%%%%%%%%%%%%%%%%%%%%%%%%%%%%%%%%%%%%%%%%%%%%%

However, for an anisotropic, stable and relativistic dynamical system $\Gamma>\frac{4}{3}$~\cite{r38,Heinzmann1975,Hillebrandt1976}, since positive anisotropy factor may slow down the growth of instability. The relativistic correction to the adiabatic index $\Gamma$ could introduce some instabilities inside the star \cite{R134,R135}. In this connection, Moustakidis \cite{R136} was proposed a more strict condition on the adiabatic index $\Gamma$ to solve this issue and achive the stable steallr structure. This condition leads the existence of a critical value for the adiabatic index $\Gamma$, denoted by $\Gamma_{crit}$, which depends on the amplitude of the Lagrangian displacement from equilibrium and the compactness factor $u=M/R$ (where $M$ and $R$ being the total mass and radius of the spherical system). Specifically in the present situation, the critical relativistic adiabatic index under in $f(R,T)$ theory can be given as
\begin{eqnarray}
\Gamma_{crit} &=& \frac{4}{3} + \frac{19}{21}u. \label{eq-adi2}
\end{eqnarray}

 The variation of the adiabatic index ($\Gamma$) with respect to the radial coordinate $r/r_b$ is shown in Fig. \ref{f10} which is physically satisfactory.
 
%%%%%%%%%%%%%%%%%%%%%%%%%%%%%%%%%%%%%%%%%%%%%%%%%%%%%%%%%%%%%%%
\begin{table}
    \centering
\caption{The numerical values of physical parameters  effective central density ($\rho^{\text{eff}}_0$), effective surface density ($\rho^{\text{eff}}_s$), and effective central pressure ($p^{\text{eff}}_0$) for  $\beta = 0.004$, $\alpha=1.4$, $\gamma=-0.002$, and $Y=0.005$.  }
\begin{tabular}{cccc}
\hline
$\chi$ and $\beta$ & {$\rho^{\text{eff}}_0\times10^{13}$} &  {$\rho^{\text{eff}}_s\times10^{13}$} & {$p^{\text{eff}}_0\times10^{33}$} \\
 & ($gm/cm^3$) & ($gm/cm^3$) & ($dyne/cm^2$)\\
\hline
$\chi$ =  0.4, $\beta$ = 0	& 3.06204	& 1.34680	& 1.84404\\
$\chi$ = -0.4, $\beta$ = 0		& 3.36193	& 1.47900	& 2.02227	\\

$\chi$ =  0.4,  $\beta$ = 0.004		& 3.07614	& 1.35285	& 2.39666	\\

$\chi$ = -0.4, $\beta$ = 0.004		& 3.37324	& 1.48417	& 2.60919\\

$\chi$ =  0.4,  $\beta$ = 0.007		& 3.08677	& 1.35739	& 2.83532	\\
\hline
\end{tabular} \label{table1}
\end{table} 

\begin{table}
\centering
\caption{The numerical values of physical parameters   ($\rho^{\text{eff}}_0$), effective surface density ($\rho^{\text{eff}}_s$), and effective central pressure ($p^{\text{eff}}_0$) for  $\beta = 0.004$, $\alpha=1.4$, $\gamma=-0.002$, and $Y=0.005$.  }
\begin{tabular}{cccc}
\hline
$\chi$ and $\beta$  & {$\rho^{\text{eff}}_0\times10^{13}$} &  {$\rho^{\text{eff}}_s\times10^{13}$} & {$p^{\text{eff}}_0\times10^{33}$}  \\
 & ($gm/cm^3$) & ($gm/cm^3$) & ($dyne/cm^2$)\\
\hline
$\chi$ =  0.4, $\beta$ = 0 & 3.06204	& 1.34680	& 1.84404	 \\
$\chi$ = -0.4, $\beta$ = 0 & 3.36193	& 1.47900	& 2.02227	 \\

$\chi$ =  0.4, $\beta$ = -0.001	 	& 3.05853	& 1.34528	& 1.71146	 \\

$\chi$ = -0.4, $\beta$ = -0.001	 	& 3.35909	& 1.47771	& 1.88101	 \\

$\chi$ =  0.4, $\beta$ = -0.002	 	& 3.05502	& 1.34377	& 1.58106	 \\
\hline
\end{tabular} \label{table2}
\end{table} 

\begin{table}
\centering
\caption{The numerical values of physical parameters mass-radius ratio $\big(\frac{M}{R}\big)$, surface red-shift ($z_s$), central adiabatic index ($\Gamma_0$), and $\Gamma_{crit}$ for different $\beta$ with $\alpha=1.4$, $\gamma=-0.002$, $Y=0.005$ and $\chi=0.4$.  }
\begin{tabular}{ccccc}
\hline
 $\beta$ & {$u=\frac{M}{R}$ } & {$z_s$}&  {$\Gamma_0$}&  {$\Gamma_{crit}$} \\
\hline
  $\beta$ = 0	&	0.2	& 0.29113	& 1.81430	& 1.51434\\

  $\beta$ = 0.004	&	0.20009		& 0.29119	& 1.82829	& 1.51437\\

 $\beta$ = 0.007 &	0.20012	& 0.29124	& 1.84826	& 1.51439\\
\hline
\end{tabular} \label{table3}
\end{table} 

\begin{table}
\centering
\caption{The numerical values of physical parameters mass-radius ratio $\big(\frac{M}{R}\big)$, surface red-shift ($z_s$), central adiabatic index ($\Gamma_0$), and $\Gamma_{crit}$ for different $\beta$ with $\alpha=1.4$, $\gamma=-0.002$, $Y=0.005$ and $\chi=0.4$.  }
\begin{tabular}{ccccc}
\hline
 $\beta$ & {$u=\frac{M}{R}$ } & {$z_s$}&  {$\Gamma_0$}&  {$\Gamma_{crit}$} \\
\hline
 $\beta$ = 0	&	0.2	& 0.29113	& 1.81430	& 1.51434\\

 $\beta$ = -0.001	&	0.200055 & 0.291112	& 1.81400	& 1.514335\\

 $\beta$ = -0.002	&	0.200047	& 0.291095	& 1.81547	& 1.514328\\
\hline
\end{tabular} \label{table4}
\end{table} 
%%%%%%%%%%%%%%%%%%%%%%%%%%%%%%%%%%%%%%%%%%%%%%%%%%%%%%%%%%%%%%%

\subsection{Equilibrium condition}
Any stellar configuration will be in stable equilibrium state only if the combined effect of all the forces related to the system becomes zero. In the present case, the obtained modified form of Tolman~\cite{Tolman1939} and Oppenheimer-Volkoff (TOV)~\cite{Tolman1939,Oppenheimer1939} equation which comply with the hydrostatic equilibrium of the stellar structure can be written as
\begin{eqnarray}\label{eq88}
&&\hspace{-0.7cm} p^{\prime}_r+\frac{\eta^{\prime}}{2}\,(\rho+p_r)-\frac{2}{r}\,(p_t-p_r)-\beta\,\Big[\left(\theta^{1}_{\ 1}\right)^{\prime}-\frac{y^{\prime}}{2}\left(\theta^{0}_{\ 0}-\theta^{1}_{\ 1}\right)\nonumber\\&& \hspace{-0.7cm}-\frac{2}{r}\left(\theta^{2}_{\ 2}-\theta^{1}_{\ 1}\right)\Big]-\frac{\chi\,(3\rho^{\prime}-p^{\prime}_r-2p^{\prime}_t)}{6\,(4\,\pi+\chi)}+\frac{\beta\,g^\prime}{2}(p_r+\rho)=0.
\end{eqnarray}

The above modified TOV equation can be written in the form of different forces components as hydrostatic force $F_{h}=- \big[p^{\prime}_r-\beta\,\left(\theta^{1}_{\ 1}\right)^{\prime}$\big], gravitational gradient force $F_g= -\big[\frac{\eta^{\prime}}{2}\,(\rho+p_r)+\frac{\beta\,g^\prime}{2}(p_r+\rho)+\beta\,\frac{y^{\prime}}{2}\left(\theta^{0}_{\ 0}-\theta^{1}_{\ 1}\right)\big]$, anisotropic force $F_a=\Big[\frac{2}{r}\,(p_t-p_r)-\frac{2}{r}\left(\theta^{2}_{\ 2}-\theta^{1}_{\ 1}\right)\big]$, and coupling force due to $f(R,T)$-gravity $F_{\chi}=\frac{\chi\,(3\rho^{\prime}-p^{\prime}_r-2p^{\prime}_t)}{6\,(4\,\pi+\chi)}$ such that $F_h+F_g+F_a+F_\chi=0$.

We have shown the features of all the forces for different values of the parameters in Fig.~\ref{f11}. From this figure it can be observed that the fluid distribution possesses a stable equilibrium under the combined effect of all these forces involved in Eq. (\ref{eq88}).

\section{Discussion and conclusion}
In the present investigation we deal with the decoupling gravitational sources in $f(R,T)$ gravity under anisotropic matter distribution. However, the energy-momentum tensor $T_{\mu\nu}$ used here under the CGD is not the usual one as in Eq. (\ref{eq8}) rather for the present study we consider the matter Lagrangian to be ${L}_m=-\mathcal{P}$, where  $\mathcal{P}=-\frac{1}{3}\big(p_r+2\,p_t)$. The basis of this particular choice is not a general one as the matter Lagrangian enters explicitly in the field equations (\ref{eq6}) and different choices lead to different equations of motion. The adopted technique of the complete geometric deformation (CGD) for solving the system of equations seems a unique pathway. This technique provides a systematic approach which are as follows: (1) split the decoupled system into two subsystems, and (2) solve these system individually. A detail discussion on this aspect has been provided in the preliminary Sec. 5 of the present manuscript. Based on the above approach we have basically an extended case of previously studied all such investigations. The results of the present study is overall interesting, distinctive and satisfactory as far as physical viability is concerned.

Some salient features of the present investigation can be discussed as follows:

(i) We have shown variation of the radial deformation function $f(r)$ and temporal deformation function $h(r)$ with respect to the radial coordinate $r/r_b$ in Fig. \ref{f1}. From this figure, for $f(r)$ and $h(r)$, we note that both are negative and decreasing function $\forall$ $r\in (r,R]$ which give the scenario of point 2 in Case I.

(ii) In Fig. \ref{f2}, the metric functions $\lambda(r)=-\ln[\xi(r)+\alpha\,f(r)]$ and $\nu(r)=\eta(r)+\alpha\,h(r)$ are both increasing and positive throughout the star, which imply that the growth of $\eta(r)$ is faster than the deformation function $h(r)$. 

(iii) In Fig. \ref{f3}, the left panel shows the effective radial pressure ($p^{\text{eff}}_r$) and the right panel shows the effective tangential pressure ($p^{\text{eff}}_t$) with respect to the radial coordinate $r/r_b$. Since the constants $\chi$ and $\beta$ influence the radial and tangential pressures, therefore the following observations can be made from Fig. \ref{f3} which are: \\{\it For fixing $\chi$:} when we increase $\beta$, both $p^{\text{eff}}_r$ and $p^{\text{eff}}_t$ increase at the core of compact object;\\ {\it For fixing $\beta$:} $p^{\text{eff}}_r$ and $p^{\text{eff}}_t$ show decreasing value at core of the star when we move $\chi$ from negative to positive value. 

(iv) Variation of the effective energy density ($\rho^{\text{eff}}$) with respect to $r/r_b$ are shown in Fig. \ref{f9}. We set the same numerical values as used in Fig. \ref{f1}. Since effective energy density is $\rho^{\text{eff}}=(1+\beta)\,\rho$, therefore value of the effective energy density at the core and boundary of the stellar object will be decreasing for the decreasing value of $\beta$. Obviously values of the effective energy density of Case I is higher than the Case II at each point of the stellar model.

(v) The left panel of Fig. \ref{f5} shows the anisotropy for the CGD contribution  ($\Delta_{CGD}$) and the right panel shows the effective anisotropy ($\Delta^{\text{eff}}$) with respect to $r/r_b$. We set the same numerical values as used in Fig. \ref{f1}. One can observe from the Fig. \ref{f5}, the anisotropic contribution $\Delta_{CGD}$ due to CGD shows the negative and decreasing behavior within the stellar model but the effective anisotropy is still increasing when $\beta\ge0$, which shows that the growth of seed anisotropy ($\Delta_{FRT}$)  is faster than $\Delta_{CGD}$. On the other hand, we would like to mention here that when $\beta$ is positive then the effective anisotropic force $F_a=\frac{2\,\Delta^{\text{eff}}}{r}$ will introduce less effect to balance the system in order to achieve the hydrostatic Equilibrium near the surface.

(vi) Variation of the radial deformation function $f(r)$ and the temporal deformation function $h(r)$ with respect to the radial coordinate $r/r_b$ are depicted in Fig. \ref{f6}. For plotting of this figure, we use the numerical values of the constants as $\beta = 0.004$, $\alpha=1.4$, $\gamma=-0.002$, $\frac{M_0}{R}=0.2$, and $Y=0.005$. In this case II:  The deformation functions $f(r)$ and $h(r)$ are also negative and decreasing function $\forall$ $r\in (r,R]$, see Fig. \ref{f6} as happens in Case I.

(vii) Variation of the metric functions $e^{\lambda}$ and $e^{\nu}$ with respect to the radial coordinate $r/r_b$ are featured in Fig. \ref{f7}. For plotting of this figure, we use the same numerical values of constants as used in Fig. \ref{f1}. It can be observed that the deformation functions $f(r)$ and $h(r)$ are negative and decreasing function throughout the stellar model and both metric functions $\lambda(r)=-\ln[\xi(r)+\alpha\,f(r)]$ and $\nu(r)=\eta(r)+\alpha\,h(r)$ are increasing as well as positive $\forall$ $r\in[0,R]$. This clearly shows the growth of $\eta(r)$ is faster than the deformation function $h(r)$ in Case II also.

(viii)  In Fig. \ref{f8}, the left panel shows the effective radial pressure ($p^{\text{eff}}_r$) and the right panel shows the effective tangential pressure ($p^{\text{eff}}_t$) with respect to the radial coordinate $r/r_b$. We set the same numerical values as used in Fig. \ref{f1}. Here the constants $\chi$ and $\beta$ also influence the radial and tangential pressures ($p^{\text{eff}}_r$ and $p^{\text{eff}}_t$), and we observe the following important points from Fig. \ref{f3} as: When we fix $\beta$ and $\chi$ move from positive to negative, the values of the pressures $p^{\text{eff}}_r$ and $p^{\text{eff}}_t$ at the core of stellar object increase, but when $\beta$ decreasing then $p^{\text{eff}}_r$ and $p^{\text{eff}}_t$ at core also decrease for fixing $\chi$. On the other hand, if we compare the Case I ($\beta>0$) and Case II ($\beta<0$), the pressure at core in Case I is higher than the Case II (see Table 1 and 2 as well as Figs. \ref{f3} and \ref{f8}). 

(ix) In Fig. \ref{f9}, the  left panel shows the anisotropy for the CGD contribution  ($\Delta_{CGD}$) and the right panel shows the effective anisotropy ($\Delta^{\text{eff}}$) with respect to $r/r_b$. We set the same numerical values as used in Fig. \ref{f6}. Here we find an interesting observations that when we look at the left panel of  Fig. \ref{f9}, the anisotropic contribution $\Delta_{CGD}$ is increasing throughout the compact star model which shows that CGD approach can also introduce a stronger anisotropy within the object since the effective anisotropy  $\Delta^{\text{eff}}=\Delta_{FRT}+\Delta_{CGD}$ will be higher than seed anisotropy $\Delta_{FRT}$. Due to this, the effective anisotropic force $F_a=\frac{2\,\Delta^{\text{eff}}}{2}$ will produce stronger effects to balance the system in order to achieve the hydrostatic equilibrium near the surface.

(x) Variation of the adiabatic index ($\Gamma$) with respect to the radial coordinate $r/r_b$ are featured in Fig. \ref{f10}. Here, we set the numerical values of constants $\beta = 0.004$, $\alpha=1.4$, $\gamma=-0.002$, $\frac{M_0}{R}=0.2$, and $Y=0.005$. We have mentioned earlier that the anisotropy may improve the stability of the model. Here we can note from Fig. \ref{f10} that value of the adiabatic index $\Gamma$ for both Cases I and II is higher in the presence of gravitational decoupling. On the other hand, we notice from Fig. \ref{f11} that the adiabatic index $\Gamma$ is monotonic increasing towards the surface. Tables \ref{table3} and \ref{table4} show that the adiabatic index $\Gamma_0 >\Gamma_{crit}$ for both the Cases I and II and hence are stable against the radial adiabatic infinitesimal perturbations.

(xi) As earlier mention that anisotropic contribution $\Delta_{CGD}$ due gravitational decoupling is negative and decreasing, then anisotropic force will show less impact as compare to hydrostatic force for balancing the system of Case I which can be seen from Fig. \ref{f11} (left panel), while in Case II the anisotropic force is stronger than the hydrostatic force near the surface. On the other hand, this force in increasing throughout the star which implies that the anisotropic force in Case 2 will introduce a strong repulsive force to avoid the gravitational collapse.

As a final comment we would like to add here that though the present work is an extension of the work of Maurya et al. \cite{Maurya2020}, the obtained results widely differ from that due to particular CGD technique as can be observed from all the case studies in connection to the above descriptions on Figs. 1--11. As far as literature survey reveals that this is the first ever approach to solve filed equations in $f(R,T)$ gravity under the extended gravitational decoupling approach, where CGD has been employed and successfully generated physically viable solutions for anisotropic system.

\section*{Appendix}
\begingroup
\small
\begin{eqnarray}
&&\hspace{-0.3cm} \theta_{11}(r)= 4 \chi \gamma (1 + r^2 Y)^2 + 8 \gamma \pi (1 + r^2 Y)^2,\nonumber\\
&&\hspace{-0.3cm}\theta_{12}(r)=4 Y [3 \gamma (\chi + \chi r^2 Y)^2 +
   3 \pi (8 \gamma \pi (1 + r^2 Y)^2  - \alpha Y  (3 + r^2 Y)) + \chi (18 \gamma \pi (1 + r^2 Y)^2 - \alpha Y (3 + r^2 Y))],\nonumber\\
  &&\hspace{-0.3cm}\theta_{21}(r)= \frac{
    192 (\chi^2 + 6 \chi \pi + 8 \pi^2) (1 + 3 C r^2) (C - 
       Y) Y L_1}{(1 + C r^2)}+ \frac{
    768 C (\chi^2 + 6 \chi \pi + 8 \pi^2) r^2 (C - 
       Y) Y L_1 (1 + r^2 Y)}{(
    1 + C r^2)} \nonumber\\&&\hspace{0.8cm} + 
    96 (\chi^2 + 6 \chi \pi + 8 \pi^2) r^2 (C - 
       Y) Y \Psi_{21}(r)\nonumber
       \end{eqnarray}
       \begin{eqnarray}
&&\hspace{-0.3cm} \theta_{22}(r)= - (1 + 
       r^2 Y) \Big[96 (\chi^2 + 6 \chi \pi + 8 \pi^2) (C - 
          Y) \Psi_{21}(r) + 
       \frac{384 C (\chi^2 + 6 \chi \pi + 8 \pi^2) r^2 (C - 
          Y) \Psi_{21}(r)}{(1 + C r^2)}
          \nonumber\\&& \hspace{0.8cm} + 
       4 \alpha r^2 \Psi^2_{21}(r) + 
       96 (\chi^2 + 6 \chi \pi + 8 \pi^2) (C - 
          Y)  \Big\{30 \gamma (\chi^2 + 6 \chi \pi + 8 \pi^2) r^2 (C - Y) Y \frac{8 C r^2 Y L_1}{(1 + 
            C r^2)^2} \nonumber\\&&\hspace{0.8cm} -\frac{4 Y L_1}{1 + C r^2} +\frac{ 4 \alpha r^2 Y^2 L_1}{(1 +
             r^2 Y)^2}- \frac{2 \alpha Y L_1}{
          1 + r^2 Y} L_4 - 
          2 Y \big(3  \gamma\, (2 + r^2 Y) + L_2 +  L_3\big)\Big\}\Big],\nonumber
\end{eqnarray}
\endgroup
where
\begingroup
\small
\begin{eqnarray}
&& \hspace{-0.2cm} \Psi_{21}(r)=L_5 - \frac{4 Y L_1}{(1 + C r^2)}
      -\frac{-2 \alpha Y\,L_1 }{(
       1 + r^2 Y)} + L_4  - 
       2 Y\, \big[3 \chi^2 \gamma (2 + r^2 Y) + 
          L_2 +  L_3\,\big],\nonumber\\
&& \hspace{-0.2cm}L_ 1 = 3 C (\chi + 4\pi) - 4 (\chi + 3\pi) Y, 
~~~ L_ 2 = 6\pi [(5 - \alpha) Y + 4\gamma\pi (2 + r^2 Y)],~~ L_ 3 = 2\chi [(5 - \alpha) Y + 9\gamma\pi (2 + r^2 Y)],\nonumber\\
&& \hspace{-0.2cm}L_ 4 = 3 C (\chi + 4\pi) [(5 - \alpha) Y + 
    2\chi\gamma (2 + r^2 Y) + 4\gamma\pi (2 + r^2 Y)],~~ L_ 5 = 6\gamma (\chi^2 + 6\chi\pi + 8\pi^2) r^2 (C - Y) Y. \nonumber     \\
&& \hspace{-0.2cm} C_{11}(r_b)=3 \gamma (\chi + \chi r_b^2 Y)^2 + 3 \pi \big(8 \gamma \pi (1 + r_b^2 Y)^2 - \alpha Y (3 + r_b^2 Y)\big)  +\chi \big[18 \gamma \pi (1 + r_b^2 Y)^2 - \alpha Y (3 + r_b^2 Y)\big],\nonumber\\
&& \hspace{-0.2cm} C_{22}(r_b)=6 \pi \big[-4 - 3 r_b^2 Y + r_b^4 Y^2 +  \beta r_b^2 (-8 \gamma \pi (1 + r_b^2 Y)^2 + \alpha Y (3 + r_b^2 Y))\big] + \chi \big[-9 - 10 r_b^2 Y + 2 r_b^4 Y^2 \nonumber\\&&\hspace{0.9cm} + \beta \big(-36 \gamma \pi (r_b + r_b^3 Y)^2 + \alpha (3 + 8 r_b^2 Y + 2 r_b^4 Y^2)\big)\big]\nonumber
\end{eqnarray}
\endgroup

\section*{Acknowledgement}
SKM and SR acknowledge that this work is carried out under TRC Project (Grant No. BFP/RGP/CBS-/19/099), the Sultanate of Oman. SKM is thankful for continuous support and encouragement from the administration of University of Nizwa.

\end{document}